\documentclass[%
reprint,
superscriptaddress,
amsmath,amssymb,
aps,
prl
]{revtex4-2}

\usepackage{soul}
\usepackage{dsfont}
\usepackage{xcolor}
\usepackage{fontawesome}
\usepackage{graphicx}
\usepackage{dcolumn}
\usepackage{bm}
\usepackage{hyperref}
\hypersetup{breaklinks=true}

\definecolor{red}{rgb}{0.75,0,0}
\definecolor{blue}{rgb}{0,0,0.75}
\definecolor{green}{rgb}{0,0.5,0}

\begin{document}
	\title{Localized states in active fluids}
	
	\author{Luca Barberi}%
	\email{luca.barberi@unige.ch}
	\affiliation{Department of Biochemistry, University of Geneva, 1211 Geneva, Switzerland}
	\affiliation{Department of Theoretical Physics, University of Geneva, 1211 Geneva, Switzerland}
	\author{Karsten Kruse}%
	\email{karsten.kruse@unige.ch}
	\affiliation{Department of Biochemistry, University of Geneva, 1211 Geneva, Switzerland}
	\affiliation{Department of Theoretical Physics, University of Geneva, 1211 Geneva, Switzerland}
	\affiliation{NCCR for Chemical Biology, University of Geneva, 1211 Geneva, Switzerland}
	
	\date{\today}
	
	\begin{abstract}
	Biological active matter is typically tightly coupled to chemical reaction networks affecting its assembly-disassembly dynamics and stress generation. We show that localized states can emerge spontaneously if assembly of active matter is regulated by chemical species that are advected with 
	flows resulting from gradients in the active stress. The mechanochemical localized patterns form via a subcritical bifurcation and for parameter values for which  patterns do not exist in absence of the advective coupling. Our work identifies a generic mechanism underlying localized cellular patterns.
	\end{abstract}

	\maketitle
	
	Chemical reactions can lead to the emergence of patterns, that is, spatiotemporally structured densities of the chemical species involved~\cite{koch1994}. Active materials that transform chemical energy into mechanical work can self-organize flow patterns and shapes~\cite{marchetti2013}. Systems coupling these two forms of self-organization are widespread, notably in engineering and biology. Typically, in these systems, chemistry is considered to control the mechanical parts. Yet, researchers increasingly focus on situations where chemistry and mechanics are mutually affecting each other~\cite{dewit2020, bailles2022}. Despite the growing interest in these systems, our understanding of spontaneous mechanochemical patterns is still limited. 
	
	A particularly interesting example of a biological mechanochemical system is the actin cortex of animal cells. It is a thin active layer beneath the outer membrane and consists of actin filaments, myosin motors and other actin-binding proteins. It exhibits a variety of spatiotemporal patterns, some of which have been argued to result from self-organization during vital cellular processes like  migration~\cite{vicker2000,Weiner2007,Stankevicins2020} or development~\cite{solon2009, munjal2015, bailles2019}. The implications of mechanochemistry in the dynamics of the actin cortex are currently under intense theoretical scrutiny \cite{bois2011, chaudhuri2011, kumar2014, banerjee2017, staddon2022}. 
	
	Whereas the most studied patterns extend over the whole cell surface, the actin cortex also exhibits localized structures. Examples of the latter are isolated contractions, either transient~\cite{baird2017} or oscillatory~\cite{graessl2017}, observed in adherent cells, as well as isolated clusters of actin and signaling molecules in cancer cells~\cite{oser2009, bravo-cordero2011}.
	
	The mechanism underlying the localized structures mentioned above has not yet been theoretically addressed. Here, we argue that these structures correspond to localized states (LSs) of spatially extended dynamical systems, \emph{i.e.}, self-organized states where a background state -- here, the homogenous, isotropic cortex -- remains essentially unaffected except in a finite region of space. In integrable systems, LSs are well-known, for instance in the form of solitons, but they also occur in dissipative systems~\cite{knobloch2015}. However, they have not been reported for the actin cortex or, generally, active fluids. 
	
	Specifically, we use a continuum description to show that mechanochemistry can produce self-organized LSs in active fluids. Our description accounts for a generic activator-inhibitor circuit involved in actin assembly~\cite{Weiner2007,bement2015,graessl2017,michaud2022}, as well as for convective flows induced by gradients in the active stress~\cite{marchetti2013}.

	Consider an isotropic active fluid. Its state is given by the density $c$ of the actomyosin network. The time evolution of $c$ is captured by the continuity equation
	\begin{equation}\label{eq:actin_mass_conservation}
	 	\partial_t c + \nabla \cdot \bm{j}_{c} =  \alpha n_a - k_d c
	\end{equation}
	where $\nabla$ is the spatial gradient operator, and the current $\bm{j}_c = c\bm{v} - D_c \nabla c$. The current consists of a convective and a diffusive part, where the latter accounts for fluctuations in the system through the effective diffusion constant $D_c$. Since cortical dynamics occurs at low Reynolds number, we neglect inertial effects, such that the fluid velocity $\bm{v}$ is determined by force balance,
	\begin{align}
		\nabla \cdot \sigma &= \gamma \bm{v} \label{eq:force_balance}
		\\
		\sigma &= 2\eta \mathsf{v} +\tilde{\eta}\nabla\cdot\bm{v}~\mathds{1}+ \pi(c) \label{eq:stress_tensor}
	\end{align}
	Here, $\sigma$ denotes the stress tensor, and $\gamma$ is a constant with the dimensions of a friction coefficient that captures dissipation in the cortex resulting, for example, from friction between the actin network and the cell membrane. 
	
	The stress tensor has a viscous component, where $\eta$ and $\tilde{\eta}$, respectively, denote the shear and bulk viscosity of the active fluid, $\mathsf{v} = (1/2)[\nabla\bm{v} + (\nabla\bm{v})^T-(\nabla\cdot\bm{v}/d)\mathds{1}]$ is the traceless strain rate tensor, $d$ the number of spatial dimensions, and $\mathds{1}$ is the identity. The non-viscous component, $\pi(c) = [\pi_a(c) + \pi_p(c)]\mathds{1}$, includes active and passive terms, $\pi_a(c) = (\zeta\Delta\mu)_0 c^2$ and $\pi_p(c) = - b c^3$, respectively. We assume the stress to be contractile for small densities $c$, $(\zeta\Delta\mu)_0 > 0$, whereas for high densities the hydrostatic contribution dominates, $b>0$~\cite{joanny2013}.
	
	The remaining terms in Eq.~\eqref{eq:actin_mass_conservation} are discussed in Ref.~\cite{ecker2021} and describe the effects of actomyosin assembly and disassembly. In the actin cortex, assembly is mediated by active actin nucleators~\cite{etienne-manneville2002, muller2020}, whose density is $n_a$.
	 	
	The actin cytoskeleton and the network regulating its assembly and activity form an excitable medium~\cite{bement2015,graessl2017,michaud2022}. We account for this feature in terms of a generic activator-inhibitor model~\cite{Doubrovinski2008,ecker2021}. Explicitly,
	\begin{align}
			\partial_t n_a+ \nabla \cdot \bm{j}_{a}& = \omega_0 (1 + \omega n_a^2) n_i - (\omega_{d,0} + \omega_d c) n_a \label{eq:nucleator_mass_conservation_a}
			\\
			\partial_t n_i+ \nabla \cdot \bm{j}_{i}& = -\omega_0 (1 + \omega n_a^2) n_i + (\omega_{d,0} + \omega_d c) n_a , \label{eq:nucleator_mass_conservation_b}
	\end{align}
	where $n_i$ is the distribution of inactive nucleators. In addition to diffusion with respective diffusion constants $D_a$ and $D_i$, active and inactive nucleators are advected with the active fluid, such that $\bm{j}_{a} = n_a \bm{v} - D_a \nabla n_a$ and analogously for the inactive nucleators. We take $D_i\gg D_a$, because, in cells, inactive nucleators are cytosolic, whereas active ones are membrane-bound \cite{etienne-manneville2002}. All terms on the right hand side of Eqs.~\eqref{eq:nucleator_mass_conservation_a} and~\eqref{eq:nucleator_mass_conservation_b} are discussed in Ref.~\cite{ecker2021}, except for the spontaneous inactivation rate $\omega_{d,0}$. Note that they conserve the total average nucleator density, $\overline{n} = (1/\mathcal{V})\int_\mathcal{V} (n_a + n_i) d\mathcal{V}$, where $\mathcal{V}$ is the system volume. Below, we consider periodic boundary conditions. No-flux boundary conditions do not change our results.
	
	We first consider the case of one spatial dimension, $d=1$, with $x \in [-l/2, l/2]$. For simplicity, we assume that $\eta = \tilde{\eta}$. We introduce non-dimensional variables, where the unit length is $\lambda = \sqrt{\eta/\gamma}$, the typical distance over which the velocity field decays due to viscosity and friction, and the unit time is $\tau = \lambda^2/D_{i}$, the typical time taken by an inactive nucleator to diffuse a distance $\lambda$. Furthermore, densities are scaled by $\overline{n}$. We denote non-dimensional quantities with capital letters: $X = x/\lambda$, $T = t/\tau$, $C = c/\bar{n}$, $N_{a (i)} = n_{a (i)} / \bar{n}$ and so on, Supplementary Material (SM) Table~II.
	
	We first consider the case $\Omega_{d,0} = 0$. In this case, Eqs.~\eqref{eq:actin_mass_conservation}--\eqref{eq:nucleator_mass_conservation_b} have a unique homogenous steady state (H)~\cite{ecker2021}. A linear stability analysis shows that this state can become unstable in favor of heterogeneous states that span the whole system, through finite wavelength stationary or oscillatory (type $\mathrm{I_s}$ or type $\mathrm{I_o}$~\cite{cross1993}) instabilities, SM Sec.~I. Note that the conservation of $\bar{n}$ implies a neutrally-stable (large-scale) mode of the dynamics, SM Sec. I. We obtain heterogenous states by solving Eqs.~\eqref{eq:actin_mass_conservation}--\eqref{eq:nucleator_mass_conservation_b} numerically. To this end, we use a custom Julia \cite{bezanson2017} code, available online~\footnote{\url{https://github.com/lucabrb/Barberi-Kruse-2022}}.

	\begin{figure}
	\includegraphics[width=\columnwidth]{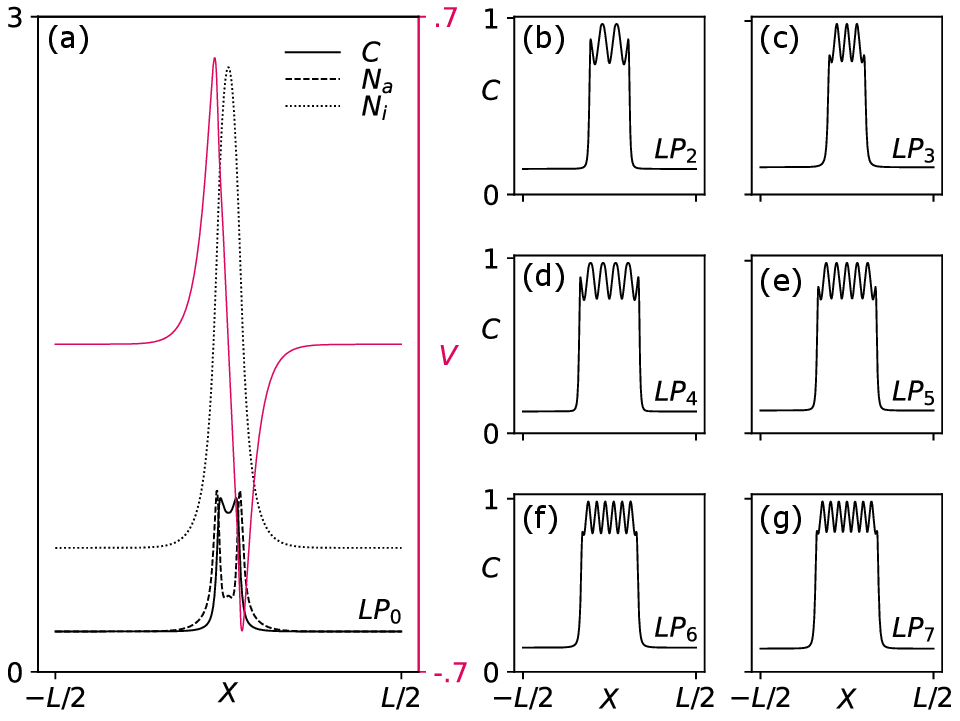}
	\caption{\label{fig:LSS} \textbf{Localized Patterns (LPs)}. a) Density (black lines) and velocity (pink line) profiles of a LP without internal peaks, LP$_0$. b--g) Multi-peaked LP$_i$, where $i = 2, \dots, 7$ is the number of internal peaks. Parameter values as in SM Table II, with $\Omega_{d,0} = 0$, $\Omega_d = 10$, $Z = 15$ and $\Omega = 6$ (a), $10$ (b, c), $14$ (d -- f), $15.5$ (g).}
	\end{figure}
	
	In addition to patterns spanning the whole system, our numerical solutions reveal a rich variety of stable localized patterns (LPs). LPs are a specific class of LSs that exhibit some internal structure and typically come in groups of related and (partially) coexisting states. In our system, LPs feature an increased active fluid and nucleator density in a confined region, Fig.~\ref{fig:LSS}a. Outside this region, the densities rapidly decay to some limiting values. With increasing system size, the high density profiles converge and the densities outside approach H, SM Fig.~3. Asymptotic convergence and lateral decay to H classifies these states as localized~\cite{champneys1998, knobloch2015}. The parameter region where LPs exist changes with increasing system size, but eventually converges to a domain of finite measure \cite{jacono2011, knobloch2016}.

	The corresponding velocity profile indicates a constant flow into the high density region, Fig.~\ref{fig:LSS}a. In addition, there is net assembly at the borders of the high-density region, SM Fig.~4. These processes are compensated by diffusive outflux from and disassembly in the center, SM Fig.~4.
	
	Additional LPs exhibit internal patterns with different numbers of density maxima (peaks), Fig.~\ref{fig:LSS}b--g. The internal patterns can be understood as a consequence of an instability of H. Here, the homogenous state H is the one of Eqs.~\eqref{eq:actin_mass_conservation}--\eqref{eq:nucleator_mass_conservation_b} with the total nucleator density $\bar{n}$ equal to that in the high density region of the states LP$_2$ to LP$_7$ and all other parameters unchanged, SM Sec.~I. Even though the linear stability analysis indicates that the system is far away from the bifurcation, the characteristic length of the LP in the high density region falls well into the interval of linearly unstable modes.
	
	The velocity profile indicates a permanent convective outflow from the internal peaks, SM Fig.~5. This enables a repulsive hydrodynamic interaction between neighboring peaks, which stabilizes LPs~\footnote{Convective outflow from the peaks is allowed by the non-monotonic dependency of $\pi(c)$ on $c$. If $\pi(c)$ has a monotonic dependence on $c$, like in Ref.~\cite{bois2011}, density peaks always generate convective inflow at their sides. In that case, the hydrodynamic interaction between neighboring peaks is attractive, which promotes their coalescence and destabilizes LPs.}. Active fluid peaks are chemically maintained by nucleation through corresponding internal peaks of active nucleator density, SM Fig.~5. These in turn are maintained by a permanent diffusive inflow of inactive nucleators that locally activate, SM Fig.~5. This structure is in line with the linear stability analysis and confirms that LPs are mechanochemical in nature instead of resulting from a chemical instability.
	
	\begin{figure}[b!]
		\includegraphics[width=\columnwidth]{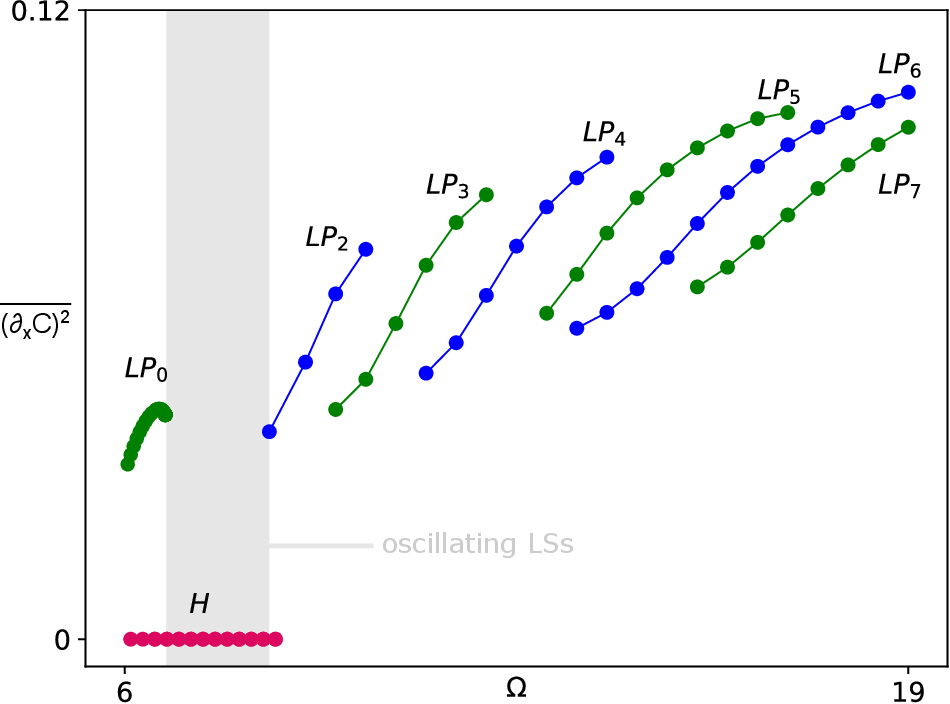}
		\caption{\label{fig:bifdiag} \textbf{Slanted snaking}. Bifurcation diagram of localized patterns (LPs). Unstable branches are not shown. In the gray-colored region, LP$_0$ loses stability to oscillatory localized states. $\overline{\left(\partial_x C\right)^2} = (1/L) \int_{-L/2}^{L/2} \left(\partial_x C\right)^2 dx$. Parameters and labels as in Fig.~\ref{fig:LSS}.}
	\end{figure} 
	
	Equations~\eqref{eq:actin_mass_conservation}--\eqref{eq:nucleator_mass_conservation_b} belong to the broad class of localized-pattern forming systems exhibiting snaking, with examples in chemistry, hydrodynamics and optics~\cite{knobloch2015, knobloch2016}. Indeed, the bifurcation diagram of LPs, Fig.~\ref{fig:bifdiag}, reveals a slanted snaking scenario~\cite{firth2007, dawes2008}, typical of systems featuring a finite-wavelength instability and a large-scale mode. In slanted snaking, LPs with an even and odd number of internal peaks form two, separate, intertwined branches (whence the name `snaking'). 
	
	 \begin{figure}
		\includegraphics[width=\columnwidth]{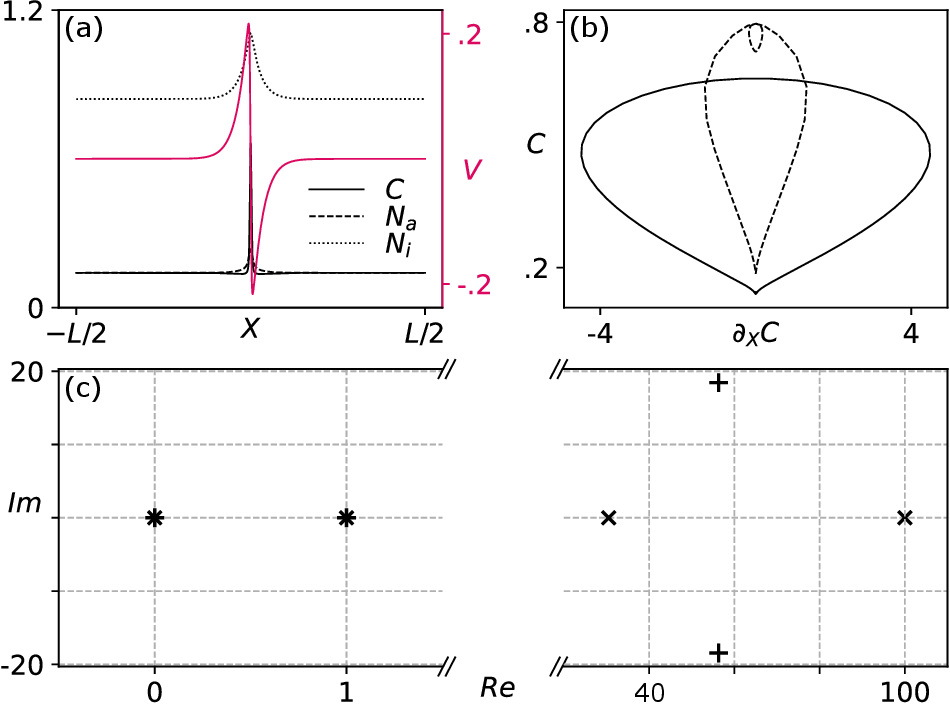}
		\caption{\label{fig:spikes} \textbf{Spikes vs. localized patterns (LPs)}. a) Profile of a spike. Parameter values as in SM Table II, with $\Omega = 0$, $\Omega_{d,0} = 3$, $\Omega_d = 0$, $Z = 20$. b) Spatial dynamics' orbits corresponding to the spike in panel (a), solid line, and LP$_0$ in Fig.~\ref{fig:LSS}a, dashed line, projected onto the $(C, dC/dX)$--plane. c) Eigenvalues of the spatial dynamics linearized around H, for parameter values corresponding to the spike in panel (a), $\times$, and LP$_0$, $+$. In either case, the eigenvalue with real part equal to 1 has multiplicity 4. To facilitate view, the horizontal axis is broken and its scale changes from left to right.}
	\end{figure}
	
	In our mechanochemical system, localization does not require nonlinear chemical reactions. Consider the linear kinetics limit of Eqs.~\eqref{eq:nucleator_mass_conservation_a} and \eqref{eq:nucleator_mass_conservation_b}, with $\Omega = \Omega_d = 0$. In this limit, the system admits LSs in the form of spikes \cite{verschueren2021} that do not have internal patterns (peaks), Fig.~\ref{fig:spikes}a. The absence of LPs in this regime shows that the internal patterns of the states in Fig.~\ref{fig:LSS}b--g and SM Fig.~5 strictly rely on cooperative nucleator activation. 

	To clarify the nature of LPs and spikes as well as the transition from one to the other, we employ spatial dynamics~\cite{kirchgassner1982, haragus2011}. That is, we consider the stationary version of Eqs.~\eqref{eq:actin_mass_conservation}--\eqref{eq:nucleator_mass_conservation_b} and interpret the spatial coordinate $X$ as a (fictitious) time. We end up with eight coupled ordinary differential equations for the effective coordinates $(C, N_a, N_i, V)$ and their effective conjugated momenta $P_C = dC/dX$, $P_{a} = dN_a/dX$, $P_i = dN_i/dX$, $P_V = dV/dX$, SM Sec.~II. The spatial dynamic system is reversible, \emph{i.e.}, symmetric for $(X, V) \to (-X, -V)$. 
	
	The homogenous stationary state H of the full dynamic equations corresponds to a fixed point of the spatial dynamics. In an infinite system, $L = \infty$, LSs map to homoclinic orbits joining H to itself in the spatial dynamics, Fig.~\ref{fig:spikes}b. These lie on the intersection of the stable and unstable manifolds of H with points on the stable manifold evolving towards H as $X\to\infty$, whereas points on the unstable manifold reach it for $X\to -\infty$. If the evolution towards and away from the fixed point is monotonic, the homoclinic orbit corresponds to a spike in the full dynamical system. Otherwise, it corresponds to an LP. This distinction is captured by the eigenvalues of the spatial dynamics linearized around H: spikes correspond to real and LPs to complex spatial eigenvalues, Fig.~\ref{fig:spikes}c. Note that eigenvalues come in complex-conjugate pairs, due to reversibility~\cite{knobloch2016}. The transition from a complex-conjugate pair (LP) to two real eigenvalues (spike) is reminiscent of a Belyakov-Devaney instability~\cite{champneys1998, verschueren2021}.
	
	\begin{figure}
		\includegraphics[width=0.94\columnwidth]{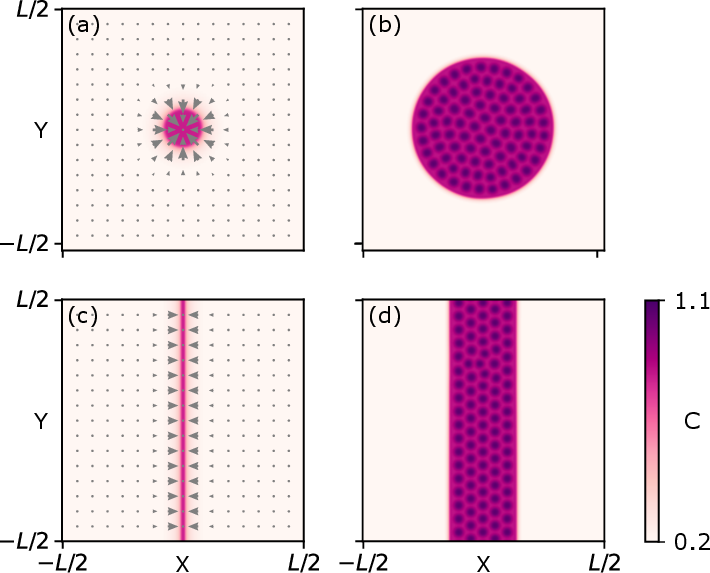}
		\caption{\label{fig:2D_LS} \textbf{Localized states in 2D.} a, b) Isotropically localized states, simple (a) and patterned (b). c, d) Anisotropically localized states, simple (c) and patterned (d). Arrows represent velocity fields (a, c). Parameters as in SM Table II, with $\Omega_{d,0} = 0$, $Z = 30$ and $\Omega = 0$ (a, c), $\Omega = 9$ (b, d). The code used to obtain these results is available online \cite{Note1}.}
	\end{figure}
 
 	LSs persist in two spatial dimensions (2D). Consider a square domain, $(x, y) \in [-l/2, l/2] \times [-l/2, l/2]$, with periodic boundaries. In 2D, we find LPs with circular, Fig.~\ref{fig:2D_LS}a,b, or band-shaped domains, Fig.~\ref{fig:2D_LS}c,d. In either case, the velocity profile indicates a constant flow into the high-density region, Fig.~\ref{fig:2D_LS}a,c. LSs with different symmetries can coexist, in which case the steady state depends on the initial conditions. Both circular and band-shaped LSs can develop internal patterns, in the form of closely packed spots, Fig.~\ref{fig:2D_LS}b,d. The velocity profile indicates a constant convective outflow from inner spots, SM Fig. 6, similar to the 1D case.	

	To conclude, motivated by the actin cortex of animal cells, we studied a generic description of an active fluid coupled to an assembly regulating module and found a spectrum of LSs. In 1D, we could identify slanted snaking to be at the origin of stationary LPs. Some LSs bifurcate into oscillating LSs, which will be presented elsewhere.  
	
	In reversible systems, homoclinic orbits to hyperbolic fixed points are structurally stable~\cite{devaney1976, champneys1998}, such that they are independent of the details of the underlying dynamic equations. In the system studied here, the fixed point H is not hyperbolic, due to the conservation of nucleators. Yet, our LSs do not rely on parameter fine tuning or the presence of chemical nonlinearities, which is consistent with structural stability.
	
	It has been argued that noise can promote transient LSs in reaction-diffusion signaling networks in absence of mechanics~\cite{hecht2010}. Like localized extracellular stimuli, these might serve as pre-patterns and trigger the mechanochemical route to stable LSs discussed above. Note that, in parameter regions where H is unstable, our system can generate LSs in absence of a localized initial condition.
	
	Due to the generic character of our approach, we refrain from a detailed comparison with experiments. Still, we want to point out some qualitative analogies between the states discussed above and cellular localized structures. In breast cancer cells, invadopodia are actin-rich protrusions that are surrounded by a ring of active Rho-C~\cite{bravo-cordero2011}. Rho-C is a small GTPase that is involved in activating actin nucleation~\cite{thomas2019}. A similar distribution is observed in the LS of Fig.~\ref{fig:LSS}a, where the active nucleator is depleted from the center of the LS, where actin density is high, and concentrates at the edges of the LSs by acquiring a two-peaked profile. Other actin-rich protrusions are known as podosomes in macrophages, where they often co-assemble into rather circular podosome clusters~\cite{hu2022}. The circular LP in Fig.~\ref{fig:2D_LS}b is reminiscent of such clusters.  
	
	Beyond invadopodia and podosome clusters, LSs could serve as programmable active stress foci to generate local membrane protrusions, like filopodia~\cite{mattila2008} and dendritic spines~\cite{bhatt2009}. As a consequence, different experimental systems are available to test the mechanism discussed in this work.
	
	\begin{acknowledgments}
		We thank Damien Brunner, Olivier Pertz, Daniel Riveline and their groups, as well as Ludovic Dumoulin, Nicolas Ecker and Oriane Foussadier for useful discussions. Numeric calculations were performed at the University of Geneva on the ``Baobab'' HPC cluster.
		This work was funded by SNF Sinergia grant CRSII5\_183550.
	\end{acknowledgments}
	
	
	\bibliography{bibliography}

\begin{thebibliography}{45}%
\makeatletter
\providecommand \@ifxundefined [1]{%
 \@ifx{#1\undefined}
}%
\providecommand \@ifnum [1]{%
 \ifnum #1\expandafter \@firstoftwo
 \else \expandafter \@secondoftwo
 \fi
}%
\providecommand \@ifx [1]{%
 \ifx #1\expandafter \@firstoftwo
 \else \expandafter \@secondoftwo
 \fi
}%
\providecommand \natexlab [1]{#1}%
\providecommand \enquote  [1]{``#1''}%
\providecommand \bibnamefont  [1]{#1}%
\providecommand \bibfnamefont [1]{#1}%
\providecommand \citenamefont [1]{#1}%
\providecommand \href@noop [0]{\@secondoftwo}%
\providecommand \href [0]{\begingroup \@sanitize@url \@href}%
\providecommand \@href[1]{\@@startlink{#1}\@@href}%
\providecommand \@@href[1]{\endgroup#1\@@endlink}%
\providecommand \@sanitize@url [0]{\catcode `\\12\catcode `\$12\catcode
  `\&12\catcode `\#12\catcode `\^12\catcode `\_12\catcode `\%12\relax}%
\providecommand \@@startlink[1]{}%
\providecommand \@@endlink[0]{}%
\providecommand \url  [0]{\begingroup\@sanitize@url \@url }%
\providecommand \@url [1]{\endgroup\@href {#1}{\urlprefix }}%
\providecommand \urlprefix  [0]{URL }%
\providecommand \Eprint [0]{\href }%
\providecommand \doibase [0]{https://doi.org/}%
\providecommand \selectlanguage [0]{\@gobble}%
\providecommand \bibinfo  [0]{\@secondoftwo}%
\providecommand \bibfield  [0]{\@secondoftwo}%
\providecommand \translation [1]{[#1]}%
\providecommand \BibitemOpen [0]{}%
\providecommand \bibitemStop [0]{}%
\providecommand \bibitemNoStop [0]{.\EOS\space}%
\providecommand \EOS [0]{\spacefactor3000\relax}%
\providecommand \BibitemShut  [1]{\csname bibitem#1\endcsname}%
\let\auto@bib@innerbib\@empty
\bibitem [{\citenamefont {Koch}\ and\ \citenamefont
  {Meinhardt}(1994)}]{koch1994}%
  \BibitemOpen
  \bibfield  {author} {\bibinfo {author} {\bibfnamefont {A.~J.}\ \bibnamefont
  {Koch}}\ and\ \bibinfo {author} {\bibfnamefont {H.}~\bibnamefont
  {Meinhardt}},\ }\bibfield  {title} {\bibinfo {title} {Biological pattern
  formation: From basic mechanisms to complex structures},\ }\href
  {https://doi.org/10.1103/RevModPhys.66.1481} {\bibfield  {journal} {\bibinfo
  {journal} {Reviews of Modern Physics}\ }\textbf {\bibinfo {volume} {66}},\
  \bibinfo {pages} {1481} (\bibinfo {year} {1994})}\BibitemShut {NoStop}%
\bibitem [{\citenamefont {Marchetti}\ \emph {et~al.}(2013)\citenamefont
  {Marchetti}, \citenamefont {Joanny}, \citenamefont {Ramaswamy}, \citenamefont
  {Liverpool}, \citenamefont {Prost}, \citenamefont {Rao},\ and\ \citenamefont
  {Simha}}]{marchetti2013}%
  \BibitemOpen
  \bibfield  {author} {\bibinfo {author} {\bibfnamefont {M.~C.}\ \bibnamefont
  {Marchetti}}, \bibinfo {author} {\bibfnamefont {J.~F.}\ \bibnamefont
  {Joanny}}, \bibinfo {author} {\bibfnamefont {S.}~\bibnamefont {Ramaswamy}},
  \bibinfo {author} {\bibfnamefont {T.~B.}\ \bibnamefont {Liverpool}}, \bibinfo
  {author} {\bibfnamefont {J.}~\bibnamefont {Prost}}, \bibinfo {author}
  {\bibfnamefont {M.}~\bibnamefont {Rao}},\ and\ \bibinfo {author}
  {\bibfnamefont {R.~A.}\ \bibnamefont {Simha}},\ }\bibfield  {title} {\bibinfo
  {title} {Hydrodynamics of soft active matter},\ }\href
  {https://doi.org/10.1103/RevModPhys.85.1143} {\bibfield  {journal} {\bibinfo
  {journal} {Reviews of Modern Physics}\ }\textbf {\bibinfo {volume} {85}},\
  \bibinfo {pages} {1143} (\bibinfo {year} {2013})}\BibitemShut {NoStop}%
\bibitem [{\citenamefont {De~Wit}(2020)}]{dewit2020}%
  \BibitemOpen
  \bibfield  {author} {\bibinfo {author} {\bibfnamefont {A.}~\bibnamefont
  {De~Wit}},\ }\bibfield  {title} {\bibinfo {title} {Chemo-{{Hydrodynamic
  Patterns}} and {{Instabilities}}},\ }\href
  {https://doi.org/10.1146/annurev-fluid-010719-060349} {\bibfield  {journal}
  {\bibinfo  {journal} {Annual Review of Fluid Mechanics}\ }\textbf {\bibinfo
  {volume} {52}},\ \bibinfo {pages} {531} (\bibinfo {year} {2020})}\BibitemShut
  {NoStop}%
\bibitem [{\citenamefont {Bailles}\ \emph {et~al.}(2022)\citenamefont
  {Bailles}, \citenamefont {Gehrels},\ and\ \citenamefont
  {Lecuit}}]{bailles2022}%
  \BibitemOpen
  \bibfield  {author} {\bibinfo {author} {\bibfnamefont {A.}~\bibnamefont
  {Bailles}}, \bibinfo {author} {\bibfnamefont {E.~W.}\ \bibnamefont
  {Gehrels}},\ and\ \bibinfo {author} {\bibfnamefont {T.}~\bibnamefont
  {Lecuit}},\ }\bibfield  {title} {\bibinfo {title} {Mechanochemical
  {{Principles}} of {{Spatial}} and {{Temporal Patterns}} in {{Cells}} and
  {{Tissues}}},\ }\href {https://doi.org/10.1146/annurev-cellbio-120420-095337}
  {\bibfield  {journal} {\bibinfo  {journal} {Annual Review of Cell and
  Developmental Biology}\ }\textbf {\bibinfo {volume} {38}},\ \bibinfo {pages}
  {4} (\bibinfo {year} {2022})}\BibitemShut {NoStop}%
\bibitem [{\citenamefont {Vicker}(2000)}]{vicker2000}%
  \BibitemOpen
  \bibfield  {author} {\bibinfo {author} {\bibfnamefont {M.~G.}\ \bibnamefont
  {Vicker}},\ }\bibfield  {title} {\bibinfo {title} {Reaction\textendash
  diffusion waves of actin filament polymerization/depolymerization in
  {{Dictyostelium}} pseudopodium extension and cell locomotion},\ }\href
  {https://doi.org/10.1016/S0301-4622(99)00146-5} {\bibfield  {journal}
  {\bibinfo  {journal} {Biophysical Chemistry}\ }\textbf {\bibinfo {volume}
  {84}},\ \bibinfo {pages} {87} (\bibinfo {year} {2000})}\BibitemShut {NoStop}%
\bibitem [{\citenamefont {Weiner}\ \emph {et~al.}(2007)\citenamefont {Weiner},
  \citenamefont {Marganski}, \citenamefont {Wu}, \citenamefont {Altschuler},\
  and\ \citenamefont {Kirschner}}]{Weiner2007}%
  \BibitemOpen
  \bibfield  {author} {\bibinfo {author} {\bibfnamefont {O.~D.}\ \bibnamefont
  {Weiner}}, \bibinfo {author} {\bibfnamefont {W.~A.}\ \bibnamefont
  {Marganski}}, \bibinfo {author} {\bibfnamefont {L.~F.}\ \bibnamefont {Wu}},
  \bibinfo {author} {\bibfnamefont {S.~J.}\ \bibnamefont {Altschuler}},\ and\
  \bibinfo {author} {\bibfnamefont {M.~W.}\ \bibnamefont {Kirschner}},\
  }\bibfield  {title} {\bibinfo {title} {An actin-based wave generator
  organizes cell motility},\ }\href
  {https://doi.org/10.1371/journal.pbio.0050221} {\bibfield  {journal}
  {\bibinfo  {journal} {PLoS Biology}\ }\textbf {\bibinfo {volume} {5}},\
  \bibinfo {pages} {2053} (\bibinfo {year} {2007})}\BibitemShut {NoStop}%
\bibitem [{\citenamefont {Stankevicins}\ \emph {et~al.}(2020)\citenamefont
  {Stankevicins}, \citenamefont {Ecker}, \citenamefont {Terriac}, \citenamefont
  {Maiuri}, \citenamefont {Schoppmeyer}, \citenamefont {Vargas}, \citenamefont
  {{Lennon-Dum{\'e}nil}}, \citenamefont {Piel}, \citenamefont {Qu},
  \citenamefont {Hoth}, \citenamefont {Kruse},\ and\ \citenamefont
  {Lautenschl{\"a}ger}}]{Stankevicins2020}%
  \BibitemOpen
  \bibfield  {author} {\bibinfo {author} {\bibfnamefont {L.}~\bibnamefont
  {Stankevicins}}, \bibinfo {author} {\bibfnamefont {N.}~\bibnamefont {Ecker}},
  \bibinfo {author} {\bibfnamefont {E.}~\bibnamefont {Terriac}}, \bibinfo
  {author} {\bibfnamefont {P.}~\bibnamefont {Maiuri}}, \bibinfo {author}
  {\bibfnamefont {R.}~\bibnamefont {Schoppmeyer}}, \bibinfo {author}
  {\bibfnamefont {P.}~\bibnamefont {Vargas}}, \bibinfo {author} {\bibfnamefont
  {A.-M.}\ \bibnamefont {{Lennon-Dum{\'e}nil}}}, \bibinfo {author}
  {\bibfnamefont {M.}~\bibnamefont {Piel}}, \bibinfo {author} {\bibfnamefont
  {B.}~\bibnamefont {Qu}}, \bibinfo {author} {\bibfnamefont {M.}~\bibnamefont
  {Hoth}}, \bibinfo {author} {\bibfnamefont {K.}~\bibnamefont {Kruse}},\ and\
  \bibinfo {author} {\bibfnamefont {F.}~\bibnamefont {Lautenschl{\"a}ger}},\
  }\bibfield  {title} {\bibinfo {title} {Deterministic actin waves as
  generators of cell polarization cues},\ }\href
  {https://doi.org/10.1073/pnas.1907845117} {\bibfield  {journal} {\bibinfo
  {journal} {Proceedings of the National Academy of Sciences}\ }\textbf
  {\bibinfo {volume} {117}},\ \bibinfo {pages} {826} (\bibinfo {year}
  {2020})}\BibitemShut {NoStop}%
\bibitem [{\citenamefont {Solon}\ \emph {et~al.}(2009)\citenamefont {Solon},
  \citenamefont {{Kaya-{\c C}opur}}, \citenamefont {Colombelli},\ and\
  \citenamefont {Brunner}}]{solon2009}%
  \BibitemOpen
  \bibfield  {author} {\bibinfo {author} {\bibfnamefont {J.}~\bibnamefont
  {Solon}}, \bibinfo {author} {\bibfnamefont {A.}~\bibnamefont {{Kaya-{\c
  C}opur}}}, \bibinfo {author} {\bibfnamefont {J.}~\bibnamefont {Colombelli}},\
  and\ \bibinfo {author} {\bibfnamefont {D.}~\bibnamefont {Brunner}},\
  }\bibfield  {title} {\bibinfo {title} {Pulsed {{Forces Timed}} by a
  {{Ratchet-like Mechanism Drive Directed Tissue Movement}} during {{Dorsal
  Closure}}},\ }\href {https://doi.org/10.1016/j.cell.2009.03.050} {\bibfield
  {journal} {\bibinfo  {journal} {Cell}\ }\textbf {\bibinfo {volume} {137}},\
  \bibinfo {pages} {1331} (\bibinfo {year} {2009})}\BibitemShut {NoStop}%
\bibitem [{\citenamefont {Munjal}\ \emph {et~al.}(2015)\citenamefont {Munjal},
  \citenamefont {Philippe}, \citenamefont {Munro},\ and\ \citenamefont
  {Lecuit}}]{munjal2015}%
  \BibitemOpen
  \bibfield  {author} {\bibinfo {author} {\bibfnamefont {A.}~\bibnamefont
  {Munjal}}, \bibinfo {author} {\bibfnamefont {J.-M.}\ \bibnamefont
  {Philippe}}, \bibinfo {author} {\bibfnamefont {E.}~\bibnamefont {Munro}},\
  and\ \bibinfo {author} {\bibfnamefont {T.}~\bibnamefont {Lecuit}},\
  }\bibfield  {title} {\bibinfo {title} {A self-organized biomechanical network
  drives shape changes during tissue morphogenesis},\ }\href
  {https://doi.org/10.1038/nature14603} {\bibfield  {journal} {\bibinfo
  {journal} {Nature}\ }\textbf {\bibinfo {volume} {524}},\ \bibinfo {pages}
  {351} (\bibinfo {year} {2015})}\BibitemShut {NoStop}%
\bibitem [{\citenamefont {Bailles}\ \emph {et~al.}(2019)\citenamefont
  {Bailles}, \citenamefont {Collinet}, \citenamefont {Philippe}, \citenamefont
  {Lenne}, \citenamefont {Munro},\ and\ \citenamefont {Lecuit}}]{bailles2019}%
  \BibitemOpen
  \bibfield  {author} {\bibinfo {author} {\bibfnamefont {A.}~\bibnamefont
  {Bailles}}, \bibinfo {author} {\bibfnamefont {C.}~\bibnamefont {Collinet}},
  \bibinfo {author} {\bibfnamefont {J.-M.}\ \bibnamefont {Philippe}}, \bibinfo
  {author} {\bibfnamefont {P.-F.}\ \bibnamefont {Lenne}}, \bibinfo {author}
  {\bibfnamefont {E.}~\bibnamefont {Munro}},\ and\ \bibinfo {author}
  {\bibfnamefont {T.}~\bibnamefont {Lecuit}},\ }\bibfield  {title} {\bibinfo
  {title} {Genetic induction and mechanochemical propagation of a morphogenetic
  wave},\ }\href {https://doi.org/10.1038/s41586-019-1492-9} {\bibfield
  {journal} {\bibinfo  {journal} {Nature}\ }\textbf {\bibinfo {volume} {572}},\
  \bibinfo {pages} {467} (\bibinfo {year} {2019})}\BibitemShut {NoStop}%
\bibitem [{\citenamefont {Bois}\ \emph {et~al.}(2011)\citenamefont {Bois},
  \citenamefont {J{\"u}licher},\ and\ \citenamefont {Grill}}]{bois2011}%
  \BibitemOpen
  \bibfield  {author} {\bibinfo {author} {\bibfnamefont {J.~S.}\ \bibnamefont
  {Bois}}, \bibinfo {author} {\bibfnamefont {F.}~\bibnamefont {J{\"u}licher}},\
  and\ \bibinfo {author} {\bibfnamefont {S.~W.}\ \bibnamefont {Grill}},\
  }\bibfield  {title} {\bibinfo {title} {Pattern {{Formation}} in {{Active
  Fluids}}},\ }\href {https://doi.org/10.1103/PhysRevLett.106.028103}
  {\bibfield  {journal} {\bibinfo  {journal} {Physical Review Letters}\
  }\textbf {\bibinfo {volume} {106}},\ \bibinfo {pages} {028103} (\bibinfo
  {year} {2011})}\BibitemShut {NoStop}%
\bibitem [{\citenamefont {Chaudhuri}\ \emph {et~al.}(2011)\citenamefont
  {Chaudhuri}, \citenamefont {Bhattacharya}, \citenamefont {Gowrishankar},
  \citenamefont {Mayor},\ and\ \citenamefont {Rao}}]{chaudhuri2011}%
  \BibitemOpen
  \bibfield  {author} {\bibinfo {author} {\bibfnamefont {A.}~\bibnamefont
  {Chaudhuri}}, \bibinfo {author} {\bibfnamefont {B.}~\bibnamefont
  {Bhattacharya}}, \bibinfo {author} {\bibfnamefont {K.}~\bibnamefont
  {Gowrishankar}}, \bibinfo {author} {\bibfnamefont {S.}~\bibnamefont
  {Mayor}},\ and\ \bibinfo {author} {\bibfnamefont {M.}~\bibnamefont {Rao}},\
  }\bibfield  {title} {\bibinfo {title} {Spatiotemporal regulation of chemical
  reactions by active cytoskeletal remodeling},\ }\href
  {https://doi.org/10.1073/pnas.1100007108} {\bibfield  {journal} {\bibinfo
  {journal} {Proceedings of the National Academy of Sciences}\ }\textbf
  {\bibinfo {volume} {108}},\ \bibinfo {pages} {14825} (\bibinfo {year}
  {2011})}\BibitemShut {NoStop}%
\bibitem [{\citenamefont {Kumar}\ \emph {et~al.}(2014)\citenamefont {Kumar},
  \citenamefont {Bois}, \citenamefont {J{\"u}licher},\ and\ \citenamefont
  {Grill}}]{kumar2014}%
  \BibitemOpen
  \bibfield  {author} {\bibinfo {author} {\bibfnamefont {K.~V.}\ \bibnamefont
  {Kumar}}, \bibinfo {author} {\bibfnamefont {J.~S.}\ \bibnamefont {Bois}},
  \bibinfo {author} {\bibfnamefont {F.}~\bibnamefont {J{\"u}licher}},\ and\
  \bibinfo {author} {\bibfnamefont {S.~W.}\ \bibnamefont {Grill}},\ }\bibfield
  {title} {\bibinfo {title} {Pulsatory {{Patterns}} in {{Active Fluids}}},\
  }\href {https://doi.org/10.1103/PhysRevLett.112.208101} {\bibfield  {journal}
  {\bibinfo  {journal} {Physical Review Letters}\ }\textbf {\bibinfo {volume}
  {112}},\ \bibinfo {pages} {208101} (\bibinfo {year} {2014})}\BibitemShut
  {NoStop}%
\bibitem [{\citenamefont {Banerjee}\ \emph {et~al.}(2017)\citenamefont
  {Banerjee}, \citenamefont {Munjal}, \citenamefont {Lecuit},\ and\
  \citenamefont {Rao}}]{banerjee2017}%
  \BibitemOpen
  \bibfield  {author} {\bibinfo {author} {\bibfnamefont {D.~S.}\ \bibnamefont
  {Banerjee}}, \bibinfo {author} {\bibfnamefont {A.}~\bibnamefont {Munjal}},
  \bibinfo {author} {\bibfnamefont {T.}~\bibnamefont {Lecuit}},\ and\ \bibinfo
  {author} {\bibfnamefont {M.}~\bibnamefont {Rao}},\ }\bibfield  {title}
  {\bibinfo {title} {Actomyosin pulsation and flows in an active elastomer with
  turnover and network remodeling},\ }\href
  {https://doi.org/10.1038/s41467-017-01130-1} {\bibfield  {journal} {\bibinfo
  {journal} {Nature Communications}\ }\textbf {\bibinfo {volume} {8}},\
  \bibinfo {pages} {1121} (\bibinfo {year} {2017})}\BibitemShut {NoStop}%
\bibitem [{\citenamefont {Staddon}\ \emph {et~al.}(2022)\citenamefont
  {Staddon}, \citenamefont {Munro},\ and\ \citenamefont
  {Banerjee}}]{staddon2022}%
  \BibitemOpen
  \bibfield  {author} {\bibinfo {author} {\bibfnamefont {M.~F.}\ \bibnamefont
  {Staddon}}, \bibinfo {author} {\bibfnamefont {E.~M.}\ \bibnamefont {Munro}},\
  and\ \bibinfo {author} {\bibfnamefont {S.}~\bibnamefont {Banerjee}},\
  }\bibfield  {title} {\bibinfo {title} {Pulsatile contractions and pattern
  formation in excitable actomyosin cortex},\ }\href
  {https://doi.org/10.1371/journal.pcbi.1009981} {\bibfield  {journal}
  {\bibinfo  {journal} {PLOS Computational Biology}\ }\textbf {\bibinfo
  {volume} {18}},\ \bibinfo {pages} {e1009981} (\bibinfo {year}
  {2022})}\BibitemShut {NoStop}%
\bibitem [{\citenamefont {Baird}\ \emph {et~al.}(2017)\citenamefont {Baird},
  \citenamefont {Billington}, \citenamefont {Wang}, \citenamefont {Adelstein},
  \citenamefont {Sellers}, \citenamefont {Fischer},\ and\ \citenamefont
  {Waterman}}]{baird2017}%
  \BibitemOpen
  \bibfield  {author} {\bibinfo {author} {\bibfnamefont {M.~A.}\ \bibnamefont
  {Baird}}, \bibinfo {author} {\bibfnamefont {N.}~\bibnamefont {Billington}},
  \bibinfo {author} {\bibfnamefont {A.}~\bibnamefont {Wang}}, \bibinfo {author}
  {\bibfnamefont {R.~S.}\ \bibnamefont {Adelstein}}, \bibinfo {author}
  {\bibfnamefont {J.~R.}\ \bibnamefont {Sellers}}, \bibinfo {author}
  {\bibfnamefont {R.~S.}\ \bibnamefont {Fischer}},\ and\ \bibinfo {author}
  {\bibfnamefont {C.~M.}\ \bibnamefont {Waterman}},\ }\bibfield  {title}
  {\bibinfo {title} {Local pulsatile contractions are an intrinsic property of
  the myosin {{2A}} motor in the cortical cytoskeleton of adherent cells},\
  }\href {https://doi.org/10.1091/mbc.e16-05-0335} {\bibfield  {journal}
  {\bibinfo  {journal} {Molecular Biology of the Cell}\ }\textbf {\bibinfo
  {volume} {28}},\ \bibinfo {pages} {240} (\bibinfo {year} {2017})}\BibitemShut
  {NoStop}%
\bibitem [{\citenamefont {Graessl}\ \emph {et~al.}(2017)\citenamefont
  {Graessl}, \citenamefont {Koch}, \citenamefont {Calderon}, \citenamefont
  {Kamps}, \citenamefont {Banerjee}, \citenamefont {Mazel}, \citenamefont
  {Schulze}, \citenamefont {Jungkurth}, \citenamefont {Patwardhan},
  \citenamefont {Solouk}, \citenamefont {Hampe}, \citenamefont {Hoffmann},
  \citenamefont {Dehmelt},\ and\ \citenamefont {Nalbant}}]{graessl2017}%
  \BibitemOpen
  \bibfield  {author} {\bibinfo {author} {\bibfnamefont {M.}~\bibnamefont
  {Graessl}}, \bibinfo {author} {\bibfnamefont {J.}~\bibnamefont {Koch}},
  \bibinfo {author} {\bibfnamefont {A.}~\bibnamefont {Calderon}}, \bibinfo
  {author} {\bibfnamefont {D.}~\bibnamefont {Kamps}}, \bibinfo {author}
  {\bibfnamefont {S.}~\bibnamefont {Banerjee}}, \bibinfo {author}
  {\bibfnamefont {T.}~\bibnamefont {Mazel}}, \bibinfo {author} {\bibfnamefont
  {N.}~\bibnamefont {Schulze}}, \bibinfo {author} {\bibfnamefont {J.~K.}\
  \bibnamefont {Jungkurth}}, \bibinfo {author} {\bibfnamefont {R.}~\bibnamefont
  {Patwardhan}}, \bibinfo {author} {\bibfnamefont {D.}~\bibnamefont {Solouk}},
  \bibinfo {author} {\bibfnamefont {N.}~\bibnamefont {Hampe}}, \bibinfo
  {author} {\bibfnamefont {B.}~\bibnamefont {Hoffmann}}, \bibinfo {author}
  {\bibfnamefont {L.}~\bibnamefont {Dehmelt}},\ and\ \bibinfo {author}
  {\bibfnamefont {P.}~\bibnamefont {Nalbant}},\ }\bibfield  {title} {\bibinfo
  {title} {An excitable {{Rho GTPase}} signaling network generates dynamic
  subcellular contraction patterns},\ }\href
  {https://doi.org/10.1083/jcb.201706052} {\bibfield  {journal} {\bibinfo
  {journal} {Journal of Cell Biology}\ }\textbf {\bibinfo {volume} {216}},\
  \bibinfo {pages} {4271} (\bibinfo {year} {2017})}\BibitemShut {NoStop}%
\bibitem [{\citenamefont {Oser}\ \emph {et~al.}(2009)\citenamefont {Oser},
  \citenamefont {Yamaguchi}, \citenamefont {Mader}, \citenamefont
  {{Bravo-Cordero}}, \citenamefont {Arias}, \citenamefont {Chen}, \citenamefont
  {DesMarais}, \citenamefont {{van Rheenen}}, \citenamefont {Koleske},\ and\
  \citenamefont {Condeelis}}]{oser2009}%
  \BibitemOpen
  \bibfield  {author} {\bibinfo {author} {\bibfnamefont {M.}~\bibnamefont
  {Oser}}, \bibinfo {author} {\bibfnamefont {H.}~\bibnamefont {Yamaguchi}},
  \bibinfo {author} {\bibfnamefont {C.~C.}\ \bibnamefont {Mader}}, \bibinfo
  {author} {\bibfnamefont {J.}~\bibnamefont {{Bravo-Cordero}}}, \bibinfo
  {author} {\bibfnamefont {M.}~\bibnamefont {Arias}}, \bibinfo {author}
  {\bibfnamefont {X.}~\bibnamefont {Chen}}, \bibinfo {author} {\bibfnamefont
  {V.}~\bibnamefont {DesMarais}}, \bibinfo {author} {\bibfnamefont
  {J.}~\bibnamefont {{van Rheenen}}}, \bibinfo {author} {\bibfnamefont {A.~J.}\
  \bibnamefont {Koleske}},\ and\ \bibinfo {author} {\bibfnamefont
  {J.}~\bibnamefont {Condeelis}},\ }\bibfield  {title} {\bibinfo {title}
  {Cortactin regulates cofilin and {{N-WASp}} activities to control the stages
  of invadopodium assembly and maturation},\ }\href
  {https://doi.org/10.1083/jcb.200812176} {\bibfield  {journal} {\bibinfo
  {journal} {Journal of Cell Biology}\ }\textbf {\bibinfo {volume} {186}},\
  \bibinfo {pages} {571} (\bibinfo {year} {2009})}\BibitemShut {NoStop}%
\bibitem [{\citenamefont {{Bravo-Cordero}}\ \emph {et~al.}(2011)\citenamefont
  {{Bravo-Cordero}}, \citenamefont {Oser}, \citenamefont {Chen}, \citenamefont
  {Eddy}, \citenamefont {Hodgson},\ and\ \citenamefont
  {Condeelis}}]{bravo-cordero2011}%
  \BibitemOpen
  \bibfield  {author} {\bibinfo {author} {\bibfnamefont {J.~J.}\ \bibnamefont
  {{Bravo-Cordero}}}, \bibinfo {author} {\bibfnamefont {M.}~\bibnamefont
  {Oser}}, \bibinfo {author} {\bibfnamefont {X.}~\bibnamefont {Chen}}, \bibinfo
  {author} {\bibfnamefont {R.}~\bibnamefont {Eddy}}, \bibinfo {author}
  {\bibfnamefont {L.}~\bibnamefont {Hodgson}},\ and\ \bibinfo {author}
  {\bibfnamefont {J.}~\bibnamefont {Condeelis}},\ }\bibfield  {title} {\bibinfo
  {title} {A {{Novel Spatiotemporal RhoC Activation Pathway Locally Regulates
  Cofilin Activity}} at {{Invadopodia}}},\ }\href
  {https://doi.org/10.1016/j.cub.2011.03.039} {\bibfield  {journal} {\bibinfo
  {journal} {Current Biology}\ }\textbf {\bibinfo {volume} {21}},\ \bibinfo
  {pages} {635} (\bibinfo {year} {2011})}\BibitemShut {NoStop}%
\bibitem [{\citenamefont {Knobloch}(2015)}]{knobloch2015}%
  \BibitemOpen
  \bibfield  {author} {\bibinfo {author} {\bibfnamefont {E.}~\bibnamefont
  {Knobloch}},\ }\bibfield  {title} {\bibinfo {title} {Spatial {{Localization}}
  in {{Dissipative Systems}}},\ }\href
  {https://doi.org/10.1146/annurev-conmatphys-031214-014514} {\bibfield
  {journal} {\bibinfo  {journal} {Annual Review of Condensed Matter Physics}\
  }\textbf {\bibinfo {volume} {6}},\ \bibinfo {pages} {325} (\bibinfo {year}
  {2015})}\BibitemShut {NoStop}%
\bibitem [{\citenamefont {Bement}\ \emph {et~al.}(2015)\citenamefont {Bement},
  \citenamefont {Leda}, \citenamefont {Moe}, \citenamefont {Kita},
  \citenamefont {Larson}, \citenamefont {Golding}, \citenamefont {Pfeuti},
  \citenamefont {Su}, \citenamefont {Miller}, \citenamefont {Goryachev},\ and\
  \citenamefont {{von~Dassow}}}]{bement2015}%
  \BibitemOpen
  \bibfield  {author} {\bibinfo {author} {\bibfnamefont {W.~M.}\ \bibnamefont
  {Bement}}, \bibinfo {author} {\bibfnamefont {M.}~\bibnamefont {Leda}},
  \bibinfo {author} {\bibfnamefont {A.~M.}\ \bibnamefont {Moe}}, \bibinfo
  {author} {\bibfnamefont {A.~M.}\ \bibnamefont {Kita}}, \bibinfo {author}
  {\bibfnamefont {M.~E.}\ \bibnamefont {Larson}}, \bibinfo {author}
  {\bibfnamefont {A.~E.}\ \bibnamefont {Golding}}, \bibinfo {author}
  {\bibfnamefont {C.}~\bibnamefont {Pfeuti}}, \bibinfo {author} {\bibfnamefont
  {K.-C.}\ \bibnamefont {Su}}, \bibinfo {author} {\bibfnamefont {A.~L.}\
  \bibnamefont {Miller}}, \bibinfo {author} {\bibfnamefont {A.~B.}\
  \bibnamefont {Goryachev}},\ and\ \bibinfo {author} {\bibfnamefont
  {G.}~\bibnamefont {{von~Dassow}}},\ }\bibfield  {title} {\bibinfo {title}
  {Activator\textendash inhibitor coupling between {{Rho}} signalling and actin
  assembly makes the cell cortex an excitable medium},\ }\href
  {https://doi.org/10.1038/ncb3251} {\bibfield  {journal} {\bibinfo  {journal}
  {Nature Cell Biology}\ }\textbf {\bibinfo {volume} {17}},\ \bibinfo {pages}
  {1471} (\bibinfo {year} {2015})}\BibitemShut {NoStop}%
\bibitem [{\citenamefont {Michaud}\ \emph {et~al.}(2022)\citenamefont
  {Michaud}, \citenamefont {Leda}, \citenamefont {Swider}, \citenamefont {Kim},
  \citenamefont {He}, \citenamefont {Landino}, \citenamefont {Valley},
  \citenamefont {Huisken}, \citenamefont {Goryachev}, \citenamefont {{von
  Dassow}},\ and\ \citenamefont {Bement}}]{michaud2022}%
  \BibitemOpen
  \bibfield  {author} {\bibinfo {author} {\bibfnamefont {A.}~\bibnamefont
  {Michaud}}, \bibinfo {author} {\bibfnamefont {M.}~\bibnamefont {Leda}},
  \bibinfo {author} {\bibfnamefont {Z.~T.}\ \bibnamefont {Swider}}, \bibinfo
  {author} {\bibfnamefont {S.}~\bibnamefont {Kim}}, \bibinfo {author}
  {\bibfnamefont {J.}~\bibnamefont {He}}, \bibinfo {author} {\bibfnamefont
  {J.}~\bibnamefont {Landino}}, \bibinfo {author} {\bibfnamefont {J.~R.}\
  \bibnamefont {Valley}}, \bibinfo {author} {\bibfnamefont {J.}~\bibnamefont
  {Huisken}}, \bibinfo {author} {\bibfnamefont {A.~B.}\ \bibnamefont
  {Goryachev}}, \bibinfo {author} {\bibfnamefont {G.}~\bibnamefont {{von
  Dassow}}},\ and\ \bibinfo {author} {\bibfnamefont {W.~M.}\ \bibnamefont
  {Bement}},\ }\bibfield  {title} {\bibinfo {title} {A versatile cortical
  pattern-forming circuit based on {{Rho}}, {{F-actin}}, {{Ect2}}, and
  {{RGA-3}}/4},\ }\href {https://doi.org/10.1083/jcb.202203017} {\bibfield
  {journal} {\bibinfo  {journal} {Journal of Cell Biology}\ }\textbf {\bibinfo
  {volume} {221}},\ \bibinfo {pages} {e202203017} (\bibinfo {year}
  {2022})}\BibitemShut {NoStop}%
\bibitem [{\citenamefont {Joanny}\ \emph {et~al.}(2013)\citenamefont {Joanny},
  \citenamefont {Kruse}, \citenamefont {Prost},\ and\ \citenamefont
  {Ramaswamy}}]{joanny2013}%
  \BibitemOpen
  \bibfield  {author} {\bibinfo {author} {\bibfnamefont {J.~F.}\ \bibnamefont
  {Joanny}}, \bibinfo {author} {\bibfnamefont {K.}~\bibnamefont {Kruse}},
  \bibinfo {author} {\bibfnamefont {J.}~\bibnamefont {Prost}},\ and\ \bibinfo
  {author} {\bibfnamefont {S.}~\bibnamefont {Ramaswamy}},\ }\bibfield  {title}
  {\bibinfo {title} {The actin cortex as an active wetting layer},\ }\href
  {https://doi.org/10.1140/epje/i2013-13052-9} {\bibfield  {journal} {\bibinfo
  {journal} {The European Physical Journal E}\ }\textbf {\bibinfo {volume}
  {36}},\ \bibinfo {pages} {52} (\bibinfo {year} {2013})}\BibitemShut {NoStop}%
\bibitem [{\citenamefont {Ecker}\ and\ \citenamefont
  {Kruse}(2021)}]{ecker2021}%
  \BibitemOpen
  \bibfield  {author} {\bibinfo {author} {\bibfnamefont {N.}~\bibnamefont
  {Ecker}}\ and\ \bibinfo {author} {\bibfnamefont {K.}~\bibnamefont {Kruse}},\
  }\bibfield  {title} {\bibinfo {title} {Excitable actin dynamics and amoeboid
  cell migration},\ }\href {https://doi.org/10.1371/journal.pone.0246311}
  {\bibfield  {journal} {\bibinfo  {journal} {PLOS ONE}\ }\textbf {\bibinfo
  {volume} {16}},\ \bibinfo {pages} {e0246311} (\bibinfo {year}
  {2021})}\BibitemShut {NoStop}%
\bibitem [{\citenamefont {{Etienne-Manneville}}\ and\ \citenamefont
  {Hall}(2002)}]{etienne-manneville2002}%
  \BibitemOpen
  \bibfield  {author} {\bibinfo {author} {\bibfnamefont {S.}~\bibnamefont
  {{Etienne-Manneville}}}\ and\ \bibinfo {author} {\bibfnamefont
  {A.}~\bibnamefont {Hall}},\ }\bibfield  {title} {\bibinfo {title} {Rho
  {{GTPases}} in cell biology},\ }\href {https://doi.org/10.1038/nature01148}
  {\bibfield  {journal} {\bibinfo  {journal} {Nature}\ }\textbf {\bibinfo
  {volume} {420}},\ \bibinfo {pages} {629} (\bibinfo {year}
  {2002})}\BibitemShut {NoStop}%
\bibitem [{\citenamefont {M{\"u}ller}\ \emph {et~al.}(2020)\citenamefont
  {M{\"u}ller}, \citenamefont {Rademacher}, \citenamefont {Bagshaw},
  \citenamefont {Wortmann}, \citenamefont {Barth}, \citenamefont {{van Unen}},
  \citenamefont {Alp}, \citenamefont {Giudice}, \citenamefont {Eccles},
  \citenamefont {Heinrich}, \citenamefont {{Pascual-Vargas}}, \citenamefont
  {{Sanchez-Castro}}, \citenamefont {Brandenburg}, \citenamefont {Mbamalu},
  \citenamefont {Tucholska}, \citenamefont {Spatt}, \citenamefont {Czajkowski},
  \citenamefont {Welke}, \citenamefont {Zhang}, \citenamefont {Nguyen},
  \citenamefont {Rrustemi}, \citenamefont {Trnka}, \citenamefont {Freitag},
  \citenamefont {Larsen}, \citenamefont {Popp}, \citenamefont {Mertins},
  \citenamefont {Gingras}, \citenamefont {Roth}, \citenamefont {Colwill},
  \citenamefont {Bakal}, \citenamefont {Pertz}, \citenamefont {Pawson},
  \citenamefont {Petsalaki},\ and\ \citenamefont {Rocks}}]{muller2020}%
  \BibitemOpen
  \bibfield  {author} {\bibinfo {author} {\bibfnamefont {P.~M.}\ \bibnamefont
  {M{\"u}ller}}, \bibinfo {author} {\bibfnamefont {J.}~\bibnamefont
  {Rademacher}}, \bibinfo {author} {\bibfnamefont {R.~D.}\ \bibnamefont
  {Bagshaw}}, \bibinfo {author} {\bibfnamefont {C.}~\bibnamefont {Wortmann}},
  \bibinfo {author} {\bibfnamefont {C.}~\bibnamefont {Barth}}, \bibinfo
  {author} {\bibfnamefont {J.}~\bibnamefont {{van Unen}}}, \bibinfo {author}
  {\bibfnamefont {K.~M.}\ \bibnamefont {Alp}}, \bibinfo {author} {\bibfnamefont
  {G.}~\bibnamefont {Giudice}}, \bibinfo {author} {\bibfnamefont {R.~L.}\
  \bibnamefont {Eccles}}, \bibinfo {author} {\bibfnamefont {L.~E.}\
  \bibnamefont {Heinrich}}, \bibinfo {author} {\bibfnamefont {P.}~\bibnamefont
  {{Pascual-Vargas}}}, \bibinfo {author} {\bibfnamefont {M.}~\bibnamefont
  {{Sanchez-Castro}}}, \bibinfo {author} {\bibfnamefont {L.}~\bibnamefont
  {Brandenburg}}, \bibinfo {author} {\bibfnamefont {G.}~\bibnamefont
  {Mbamalu}}, \bibinfo {author} {\bibfnamefont {M.}~\bibnamefont {Tucholska}},
  \bibinfo {author} {\bibfnamefont {L.}~\bibnamefont {Spatt}}, \bibinfo
  {author} {\bibfnamefont {M.~T.}\ \bibnamefont {Czajkowski}}, \bibinfo
  {author} {\bibfnamefont {R.-W.}\ \bibnamefont {Welke}}, \bibinfo {author}
  {\bibfnamefont {S.}~\bibnamefont {Zhang}}, \bibinfo {author} {\bibfnamefont
  {V.}~\bibnamefont {Nguyen}}, \bibinfo {author} {\bibfnamefont
  {T.}~\bibnamefont {Rrustemi}}, \bibinfo {author} {\bibfnamefont
  {P.}~\bibnamefont {Trnka}}, \bibinfo {author} {\bibfnamefont
  {K.}~\bibnamefont {Freitag}}, \bibinfo {author} {\bibfnamefont
  {B.}~\bibnamefont {Larsen}}, \bibinfo {author} {\bibfnamefont
  {O.}~\bibnamefont {Popp}}, \bibinfo {author} {\bibfnamefont {P.}~\bibnamefont
  {Mertins}}, \bibinfo {author} {\bibfnamefont {A.-C.}\ \bibnamefont
  {Gingras}}, \bibinfo {author} {\bibfnamefont {F.~P.}\ \bibnamefont {Roth}},
  \bibinfo {author} {\bibfnamefont {K.}~\bibnamefont {Colwill}}, \bibinfo
  {author} {\bibfnamefont {C.}~\bibnamefont {Bakal}}, \bibinfo {author}
  {\bibfnamefont {O.}~\bibnamefont {Pertz}}, \bibinfo {author} {\bibfnamefont
  {T.}~\bibnamefont {Pawson}}, \bibinfo {author} {\bibfnamefont
  {E.}~\bibnamefont {Petsalaki}},\ and\ \bibinfo {author} {\bibfnamefont
  {O.}~\bibnamefont {Rocks}},\ }\bibfield  {title} {\bibinfo {title} {Systems
  analysis of {{RhoGEF}} and {{RhoGAP}} regulatory proteins reveals spatially
  organized {{RAC1}} signalling from integrin adhesions},\ }\href
  {https://doi.org/10.1038/s41556-020-0488-x} {\bibfield  {journal} {\bibinfo
  {journal} {Nature Cell Biology}\ }\textbf {\bibinfo {volume} {22}},\ \bibinfo
  {pages} {498} (\bibinfo {year} {2020})}\BibitemShut {NoStop}%
\bibitem [{\citenamefont {Doubrovinski}\ and\ \citenamefont
  {Kruse}(2008)}]{Doubrovinski2008}%
  \BibitemOpen
  \bibfield  {author} {\bibinfo {author} {\bibfnamefont {K.}~\bibnamefont
  {Doubrovinski}}\ and\ \bibinfo {author} {\bibfnamefont {K.}~\bibnamefont
  {Kruse}},\ }\bibfield  {title} {\bibinfo {title} {Cytoskeletal waves in the
  absence of molecular motors},\ }\href
  {https://doi.org/10.1209/0295-5075/83/18003} {\bibfield  {journal} {\bibinfo
  {journal} {EPL (Europhysics Letters)}\ }\textbf {\bibinfo {volume} {83}},\
  \bibinfo {pages} {18003} (\bibinfo {year} {2008})}\BibitemShut {NoStop}%
\bibitem [{\citenamefont {Cross}\ and\ \citenamefont
  {Hohenberg}(1993)}]{cross1993}%
  \BibitemOpen
  \bibfield  {author} {\bibinfo {author} {\bibfnamefont {M.~C.}\ \bibnamefont
  {Cross}}\ and\ \bibinfo {author} {\bibfnamefont {P.~C.}\ \bibnamefont
  {Hohenberg}},\ }\bibfield  {title} {\bibinfo {title} {Pattern formation
  outside of equilibrium},\ }\href {https://doi.org/10.1103/RevModPhys.65.851}
  {\bibfield  {journal} {\bibinfo  {journal} {Reviews of Modern Physics}\
  }\textbf {\bibinfo {volume} {65}},\ \bibinfo {pages} {851} (\bibinfo {year}
  {1993})}\BibitemShut {NoStop}%
\bibitem [{\citenamefont {Bezanson}\ \emph {et~al.}(2017)\citenamefont
  {Bezanson}, \citenamefont {Edelman}, \citenamefont {Karpinski},\ and\
  \citenamefont {Shah}}]{bezanson2017}%
  \BibitemOpen
  \bibfield  {author} {\bibinfo {author} {\bibfnamefont {J.}~\bibnamefont
  {Bezanson}}, \bibinfo {author} {\bibfnamefont {A.}~\bibnamefont {Edelman}},
  \bibinfo {author} {\bibfnamefont {S.}~\bibnamefont {Karpinski}},\ and\
  \bibinfo {author} {\bibfnamefont {V.~B.}\ \bibnamefont {Shah}},\ }\bibfield
  {title} {\bibinfo {title} {Julia: {{A Fresh Approach}} to {{Numerical
  Computing}}},\ }\href {https://doi.org/10.1137/141000671} {\bibfield
  {journal} {\bibinfo  {journal} {SIAM Review}\ }\textbf {\bibinfo {volume}
  {59}},\ \bibinfo {pages} {65} (\bibinfo {year} {2017})}\BibitemShut {NoStop}%
\bibitem [{Note1()}]{Note1}%
  \BibitemOpen
  \bibinfo {note} {\protect \url
  {https://github.com/lucabrb/Barberi-Kruse-2022}}\BibitemShut {NoStop}%
\bibitem [{\citenamefont {Champneys}(1998)}]{champneys1998}%
  \BibitemOpen
  \bibfield  {author} {\bibinfo {author} {\bibfnamefont {A.~R.}\ \bibnamefont
  {Champneys}},\ }\bibfield  {title} {\bibinfo {title} {Homoclinic orbits in
  reversible systems and their applications in mechanics, fluids and optics},\
  }\href@noop {} {\bibfield  {journal} {\bibinfo  {journal} {Physica D}\ ,\
  \bibinfo {pages} {29}} (\bibinfo {year} {1998})}\BibitemShut {NoStop}%
\bibitem [{\citenamefont {Jacono}\ \emph {et~al.}(2011)\citenamefont {Jacono},
  \citenamefont {Bergeon},\ and\ \citenamefont {Knobloch}}]{jacono2011}%
  \BibitemOpen
  \bibfield  {author} {\bibinfo {author} {\bibfnamefont {D.~L.}\ \bibnamefont
  {Jacono}}, \bibinfo {author} {\bibfnamefont {A.}~\bibnamefont {Bergeon}},\
  and\ \bibinfo {author} {\bibfnamefont {E.}~\bibnamefont {Knobloch}},\
  }\bibfield  {title} {\bibinfo {title} {Magnetohydrodynamic convectons},\
  }\href {https://doi.org/10.1017/jfm.2011.402} {\bibfield  {journal} {\bibinfo
   {journal} {Journal of Fluid Mechanics}\ }\textbf {\bibinfo {volume} {687}},\
  \bibinfo {pages} {595} (\bibinfo {year} {2011})}\BibitemShut {NoStop}%
\bibitem [{\citenamefont {Knobloch}(2016)}]{knobloch2016}%
  \BibitemOpen
  \bibfield  {author} {\bibinfo {author} {\bibfnamefont {E.}~\bibnamefont
  {Knobloch}},\ }\bibfield  {title} {\bibinfo {title} {Localized structures and
  front propagation in systems with a conservation law},\ }\href
  {https://doi.org/10.1093/imamat/hxw029} {\bibfield  {journal} {\bibinfo
  {journal} {IMA Journal of Applied Mathematics}\ }\textbf {\bibinfo {volume}
  {81}},\ \bibinfo {pages} {457} (\bibinfo {year} {2016})}\BibitemShut
  {NoStop}%
\bibitem [{Note2()}]{Note2}%
  \BibitemOpen
  \bibinfo {note} {Convective outflow from the peaks is allowed by the
  non-monotonic dependency of $\pi (c)$ on $c$. If $\pi (c)$ has a monotonic
  dependence on $c$, like in Ref.~\cite {bois2011}, density peaks always
  generate convective inflow at their sides. In that case, the hydrodynamic
  interaction between neighboring peaks is attractive, which promotes their
  coalescence and destabilizes LPs.}\BibitemShut {Stop}%
\bibitem [{\citenamefont {Firth}\ \emph {et~al.}(2007)\citenamefont {Firth},
  \citenamefont {Columbo},\ and\ \citenamefont {Scroggie}}]{firth2007}%
  \BibitemOpen
  \bibfield  {author} {\bibinfo {author} {\bibfnamefont {W.~J.}\ \bibnamefont
  {Firth}}, \bibinfo {author} {\bibfnamefont {L.}~\bibnamefont {Columbo}},\
  and\ \bibinfo {author} {\bibfnamefont {A.~J.}\ \bibnamefont {Scroggie}},\
  }\bibfield  {title} {\bibinfo {title} {Proposed {{Resolution}} of
  {{Theory-Experiment Discrepancy}} in {{Homoclinic Snaking}}},\ }\href
  {https://doi.org/10.1103/PhysRevLett.99.104503} {\bibfield  {journal}
  {\bibinfo  {journal} {Physical Review Letters}\ }\textbf {\bibinfo {volume}
  {99}},\ \bibinfo {pages} {104503} (\bibinfo {year} {2007})}\BibitemShut
  {NoStop}%
\bibitem [{\citenamefont {Dawes}(2008)}]{dawes2008}%
  \BibitemOpen
  \bibfield  {author} {\bibinfo {author} {\bibfnamefont {J.~H.~P.}\
  \bibnamefont {Dawes}},\ }\bibfield  {title} {\bibinfo {title} {Localized
  {{Pattern Formation}} with a {{Large-Scale Mode}}: {{Slanted Snaking}}},\
  }\href {https://doi.org/10.1137/06067794X} {\bibfield  {journal} {\bibinfo
  {journal} {SIAM Journal on Applied Dynamical Systems}\ }\textbf {\bibinfo
  {volume} {7}},\ \bibinfo {pages} {186} (\bibinfo {year} {2008})}\BibitemShut
  {NoStop}%
\bibitem [{\citenamefont {Verschueren}\ and\ \citenamefont
  {Champneys}(2021)}]{verschueren2021}%
  \BibitemOpen
  \bibfield  {author} {\bibinfo {author} {\bibfnamefont {N.}~\bibnamefont
  {Verschueren}}\ and\ \bibinfo {author} {\bibfnamefont {A.~R.}\ \bibnamefont
  {Champneys}},\ }\bibfield  {title} {\bibinfo {title} {Dissecting the snake:
  {{Transition}} from localized patterns to spike solutions},\ }\href
  {https://doi.org/10.1016/j.physd.2021.132858} {\bibfield  {journal} {\bibinfo
   {journal} {Physica D: Nonlinear Phenomena}\ }\textbf {\bibinfo {volume}
  {419}},\ \bibinfo {pages} {132858} (\bibinfo {year} {2021})}\BibitemShut
  {NoStop}%
\bibitem [{\citenamefont {Kirchg{\"a}ssner}(1982)}]{kirchgassner1982}%
  \BibitemOpen
  \bibfield  {author} {\bibinfo {author} {\bibfnamefont {K.}~\bibnamefont
  {Kirchg{\"a}ssner}},\ }\bibfield  {title} {\bibinfo {title} {Wave-solutions
  of reversible systems and applications},\ }\href
  {https://doi.org/10.1016/0022-0396(82)90058-4} {\bibfield  {journal}
  {\bibinfo  {journal} {Journal of Differential Equations}\ }\textbf {\bibinfo
  {volume} {45}},\ \bibinfo {pages} {113} (\bibinfo {year} {1982})}\BibitemShut
  {NoStop}%
\bibitem [{\citenamefont {Haragus}\ and\ \citenamefont
  {Iooss}(2011)}]{haragus2011}%
  \BibitemOpen
  \bibfield  {author} {\bibinfo {author} {\bibfnamefont {M.}~\bibnamefont
  {Haragus}}\ and\ \bibinfo {author} {\bibfnamefont {G.}~\bibnamefont
  {Iooss}},\ }\href {https://doi.org/10.1007/978-0-85729-112-7} {\emph
  {\bibinfo {title} {Local {{Bifurcations}}, {{Center Manifolds}}, and {{Normal
  Forms}} in {{Infinite-Dimensional Dynamical Systems}}}}}\ (\bibinfo
  {publisher} {{Springer London}},\ \bibinfo {address} {{London}},\ \bibinfo
  {year} {2011})\BibitemShut {NoStop}%
\bibitem [{\citenamefont {Devaney}(1976)}]{devaney1976}%
  \BibitemOpen
  \bibfield  {author} {\bibinfo {author} {\bibfnamefont {R.~L.}\ \bibnamefont
  {Devaney}},\ }\bibfield  {title} {\bibinfo {title} {Reversible
  {{Diffeomorphisms}} and {{Flows}}},\ }\href {https://doi.org/10.2307/1997429}
  {\bibfield  {journal} {\bibinfo  {journal} {Transactions of the American
  Mathematical Society}\ }\textbf {\bibinfo {volume} {218}},\ \bibinfo {pages}
  {89} (\bibinfo {year} {1976})},\ \Eprint {https://arxiv.org/abs/1997429}
  {1997429} \BibitemShut {NoStop}%
\bibitem [{\citenamefont {Hecht}\ \emph {et~al.}(2010)\citenamefont {Hecht},
  \citenamefont {Kessler},\ and\ \citenamefont {Levine}}]{hecht2010}%
  \BibitemOpen
  \bibfield  {author} {\bibinfo {author} {\bibfnamefont {I.}~\bibnamefont
  {Hecht}}, \bibinfo {author} {\bibfnamefont {D.~A.}\ \bibnamefont {Kessler}},\
  and\ \bibinfo {author} {\bibfnamefont {H.}~\bibnamefont {Levine}},\
  }\bibfield  {title} {\bibinfo {title} {Transient {{Localized Patterns}} in
  {{Noise-Driven Reaction-Diffusion Systems}}},\ }\href
  {https://doi.org/10.1103/PhysRevLett.104.158301} {\bibfield  {journal}
  {\bibinfo  {journal} {Physical Review Letters}\ }\textbf {\bibinfo {volume}
  {104}},\ \bibinfo {pages} {158301} (\bibinfo {year} {2010})}\BibitemShut
  {NoStop}%
\bibitem [{\citenamefont {Thomas}\ \emph {et~al.}(2019)\citenamefont {Thomas},
  \citenamefont {Pranatharthi}, \citenamefont {Ross},\ and\ \citenamefont
  {Srivastava}}]{thomas2019}%
  \BibitemOpen
  \bibfield  {author} {\bibinfo {author} {\bibfnamefont {P.}~\bibnamefont
  {Thomas}}, \bibinfo {author} {\bibfnamefont {A.}~\bibnamefont
  {Pranatharthi}}, \bibinfo {author} {\bibfnamefont {C.}~\bibnamefont {Ross}},\
  and\ \bibinfo {author} {\bibfnamefont {S.}~\bibnamefont {Srivastava}},\
  }\bibfield  {title} {\bibinfo {title} {{{RhoC}}: A fascinating journey from a
  cytoskeletal organizer to a {{Cancer}} stem cell therapeutic target},\ }\href
  {https://doi.org/10.1186/s13046-019-1327-4} {\bibfield  {journal} {\bibinfo
  {journal} {Journal of Experimental \& Clinical Cancer Research}\ }\textbf
  {\bibinfo {volume} {38}},\ \bibinfo {pages} {328} (\bibinfo {year}
  {2019})}\BibitemShut {NoStop}%
\bibitem [{\citenamefont {Hu}\ \emph {et~al.}(2022)\citenamefont {Hu},
  \citenamefont {Zhu}, \citenamefont {Dong}, \citenamefont {Zhang},
  \citenamefont {Xing}, \citenamefont {Li}, \citenamefont {Yan}, \citenamefont
  {Zhou}, \citenamefont {Xu}, \citenamefont {Pan},\ and\ \citenamefont
  {Xu}}]{hu2022}%
  \BibitemOpen
  \bibfield  {author} {\bibinfo {author} {\bibfnamefont {F.}~\bibnamefont
  {Hu}}, \bibinfo {author} {\bibfnamefont {D.}~\bibnamefont {Zhu}}, \bibinfo
  {author} {\bibfnamefont {H.}~\bibnamefont {Dong}}, \bibinfo {author}
  {\bibfnamefont {P.}~\bibnamefont {Zhang}}, \bibinfo {author} {\bibfnamefont
  {F.}~\bibnamefont {Xing}}, \bibinfo {author} {\bibfnamefont {W.}~\bibnamefont
  {Li}}, \bibinfo {author} {\bibfnamefont {R.}~\bibnamefont {Yan}}, \bibinfo
  {author} {\bibfnamefont {J.}~\bibnamefont {Zhou}}, \bibinfo {author}
  {\bibfnamefont {K.}~\bibnamefont {Xu}}, \bibinfo {author} {\bibfnamefont
  {L.}~\bibnamefont {Pan}},\ and\ \bibinfo {author} {\bibfnamefont
  {J.}~\bibnamefont {Xu}},\ }\bibfield  {title} {\bibinfo {title}
  {Super-resolution microscopy reveals nanoscale architecture and regulation of
  podosome clusters in primary macrophages},\ }\href
  {https://doi.org/10.1016/j.isci.2022.105514} {\bibfield  {journal} {\bibinfo
  {journal} {iScience}\ }\textbf {\bibinfo {volume} {25}},\ \bibinfo {pages}
  {105514} (\bibinfo {year} {2022})}\BibitemShut {NoStop}%
\bibitem [{\citenamefont {Mattila}\ and\ \citenamefont
  {Lappalainen}(2008)}]{mattila2008}%
  \BibitemOpen
  \bibfield  {author} {\bibinfo {author} {\bibfnamefont {P.~K.}\ \bibnamefont
  {Mattila}}\ and\ \bibinfo {author} {\bibfnamefont {P.}~\bibnamefont
  {Lappalainen}},\ }\bibfield  {title} {\bibinfo {title} {Filopodia: Molecular
  architecture and cellular functions},\ }\href
  {https://doi.org/10.1038/nrm2406} {\bibfield  {journal} {\bibinfo  {journal}
  {Nature Reviews Molecular Cell Biology}\ }\textbf {\bibinfo {volume} {9}},\
  \bibinfo {pages} {446} (\bibinfo {year} {2008})}\BibitemShut {NoStop}%
\bibitem [{\citenamefont {Bhatt}\ \emph {et~al.}(2009)\citenamefont {Bhatt},
  \citenamefont {Zhang},\ and\ \citenamefont {Gan}}]{bhatt2009}%
  \BibitemOpen
  \bibfield  {author} {\bibinfo {author} {\bibfnamefont {D.~H.}\ \bibnamefont
  {Bhatt}}, \bibinfo {author} {\bibfnamefont {S.}~\bibnamefont {Zhang}},\ and\
  \bibinfo {author} {\bibfnamefont {W.-B.}\ \bibnamefont {Gan}},\ }\bibfield
  {title} {\bibinfo {title} {Dendritic {{Spine Dynamics}}},\ }\href
  {https://doi.org/10.1146/annurev.physiol.010908.163140} {\bibfield  {journal}
  {\bibinfo  {journal} {Annual Review of Physiology}\ }\textbf {\bibinfo
  {volume} {71}},\ \bibinfo {pages} {261} (\bibinfo {year} {2009})}\BibitemShut
  {NoStop}%
\end{thebibliography}%
	
\end{document}


	\title{Supplementary Material for ``Localized states in active fluids''}
	
	\author{Luca Barberi}%
	\email{luca.barberi@unige.ch}
	\affiliation{Department of Biochemistry, University of Geneva, 1211 Geneva, Switzerland}
	\affiliation{Department of Theoretical Physics, University of Geneva, 1211 Geneva, Switzerland}
	\author{Karsten Kruse}%
	\email{karsten.kruse@unige.ch}
	\affiliation{Department of Biochemistry, University of Geneva, 1211 Geneva, Switzerland}
	\affiliation{Department of Theoretical Physics, University of Geneva, 1211 Geneva, Switzerland}
	\affiliation{NCCR for Chemical Biology, University of Geneva, 1211 Geneva, Switzerland}
	
	\date{\today}

	\maketitle

	\section{Linear Stability Analysis} \label{sec:LSA}
	In this section, we analyze Eqs.~(1--5) of the main text in terms of the linear stability of their homogeneous steady state, H. In Sec.~\ref{sec:LSA_1}, we compute a stability diagram of H in terms of the non-dimensional parameters $Z$ and $\Omega$, SM Table~\ref{table:parameters}, where finite-wavelength instabilities are identified. In Sec.~\ref{sec:LSA_2}, we study the mechanism underlying the formation of localized patterns in terms of a linear stability analysis.
	
	\subsection{Linear stability analysis of the full system, in 1D} \label{sec:LSA_1}
	To start, we formulate the equations in non-dimensional units, SM Table \ref{table:parameters}. Eq.~(1) of the main text reads
	\begin{equation}
		\label{eq:actin_dynamics}
		\partial_T C + \partial_X J_c = A N_a - K_d C,
	\end{equation}
	where $J_c = V C - \mathscr{D}_c \partial_x C$, $A = \alpha\tau$. Eqs. (2--3) of the main text read
	\begin{align}
		\partial_X \Sigma &= V \label{eq:force_balance}\\
		\Sigma &= \partial_X V + \Pi(C), \label{eq:stress_field}
	\end{align}
	with $\Pi(C) = Z C^2 - B C^3$. Finally, Eqs.~(4--5) read
	\begin{align}
		\partial_T N_a + \partial_X J_a &= \Omega_0 (1 + \Omega N_a^2)N_i - (\Omega_{d,0} + \Omega_d C) N_a \label{eq:nucleator_dynamics_a}\\
		\partial_T N_i + \partial_X J_i &= -\Omega_0 (1 + \Omega N_a^2)N_i + (\Omega_{d,0} + \Omega_d C) N_a, \label{eq:nucleator_dynamics_b}
	\end{align}
	where $J_{(a, i)} = V N_{(a, i)} - \mathscr{D}_{(a, i)} \partial_x N_{(a, i)}$. 
	
	We make a change of variables by introducing $N_+ = N_a + N_i$ and $N_- = N_a - N_i$. In terms of these new variables, Eq.~\eqref{eq:actin_dynamics} and Eqs.~\eqref{eq:nucleator_dynamics_a}--\eqref{eq:nucleator_dynamics_b} read, respectively
	\begin{equation}
		\label{eq:actin_dynamics_2}
		\partial_T C + \partial_X J_c = A \frac{N_+ + N_-}{2} - K_d C,
	\end{equation}
	and
	\begin{align}
		\partial_T N_+ + \partial_X J_+ &= 0 \label{eq:nucleator_dynamics_2_a}\\
		\partial_T N_- + \partial_X J_- &= \Omega_0 \left[1 + \Omega \left(\frac{N_+ + N_-}{2}\right)^2\right](N_+ - N_-) - (\Omega_{d,0} + \Omega_d C) (N_+ + N_-), \label{eq:nucleator_dynamics_2_b}
	\end{align}
	where $J_\pm= N_\pm V -(1/2)(\mathscr{D}_a^\pm \partial_X N_+ + \mathscr{D}_a^\mp \partial_X N_-)$ and $\mathscr{D}_a^\pm = \mathscr{D}_a \pm 1$. Note that $N_+$ is a conserved quantity, which implies a neutrally stable, large-scale mode in the linear stability analysis below.
	
	Henceforth, we use the subscript ``H'' to denote quantities evaluated at the homogeneous steady state H of Eqs.~\eqref{eq:actin_dynamics}--\eqref{eq:nucleator_dynamics_b}, \textit{i.e.} $C_\mathrm{H} = A (N_a)_\mathrm{H}$, $(N_a)_\mathrm{H}$, $(N_i)_\mathrm{H} = 1 - (N_a)_\mathrm{H}$ and $V_\mathrm{H}=0$, such that $N^+_\mathrm{H} = 1$ and $N^-_\mathrm{H} = 2(N_a)_\mathrm{H} -1$. We introduce small plane wave perturbations to H by setting  
	\begin{align}
		C &= C_\mathrm{H} + \delta C e^{iKX + \Phi T} \\
		N^+ &= 1 + \delta N^+ e^{iKX + \Phi T} \\
		N^- &= N^-_\mathrm{H} + \delta N^- e^{iKX + \Phi T} \\
		V &= \delta V e^{iKX + \Phi T}
	\end{align}
	where we assume $\delta C$, $\delta N_a$, $\delta N_i$ and $\delta V$ to be of the same order $\ll 1$, and linearize Eqs.~\eqref{eq:force_balance}, \eqref{eq:actin_dynamics_2}--\eqref{eq:nucleator_dynamics_2_b} accordingly. Note that linearizing Eq.~\eqref{eq:force_balance} provides an expression of $\delta V$ in terms of $\delta C$,
	\begin{equation}
		\delta V = \frac{iK}{1 + K^2}\Pi'_\mathrm{H} \delta C,
	\end{equation}
	where $\Pi'_\mathrm{H} = 2 Z C_\mathrm{H} - 3 B C_\mathrm{H}^2$. This reduces the linearized dynamics to a $3\times3$ eigenvalue problem $M \bm{u} = \Phi \bm{u}$, where $\bm{u} = (\delta C, \delta N^+, \delta N^-)$ and the matrix
	\begin{equation}
		M = 
		\begin{pmatrix}
			M_{CC} & M_{C+} & M_{C-}\\
			M_{+C} & M_{++} & M_{+-}\\
			M_{-C} & M_{-+} & M_{--}
		\end{pmatrix}
	\end{equation}
	has entries
	\begin{align}
		M_{CC}&= \frac{K^2}{1+K^2}\Pi_\mathrm{H}' C_\mathrm{H} - \mathscr{D}_c K^2 - K_d \\
		M_{C+}&= \frac{A}{2} \\
		M_{C-}&= \frac{A}{2} \\
		M_{+C}&= \frac{K^2}{1+K^2}\Pi_\mathrm{H}' \\
		M_{++}&= -\frac{\mathscr{D}_a^+}{2}K^2 \\
		M_{+-}&= -\frac{\mathscr{D}_a^-}{2}K^2 \\
		M_{-C}&= \frac{K^2}{1+K^2}\Pi_\mathrm{H}' N^-_\mathrm{H} - (1 + N^-_\mathrm{H})\Omega_d \\
		M_{-+}&= -\frac{\mathscr{D}_a^-}{2}K^2 + \Omega_0 - \frac{1}{4}(N^-_\mathrm{H} - 3)(N^-_\mathrm{H} + 1)\Omega\Omega_0 - \Omega_{d,0} - C_\mathrm{H}\Omega_d \\
		M_{--}&= -\frac{\mathscr{D}_a^+}{2}K^2 - \frac{1}{4}\left[4 + (1+N^-_\mathrm{H})(3N^-_\mathrm{H} - 1)\Omega\right]\Omega_0 - \Omega_{d,0} - C_\mathrm{H}\Omega_d. 
	\end{align}
	Note that the second line of $M$ vanishes at $K=0$, yielding the large-scale mode mentioned above.  

	\begin{figure}
		\includegraphics[width=\columnwidth]{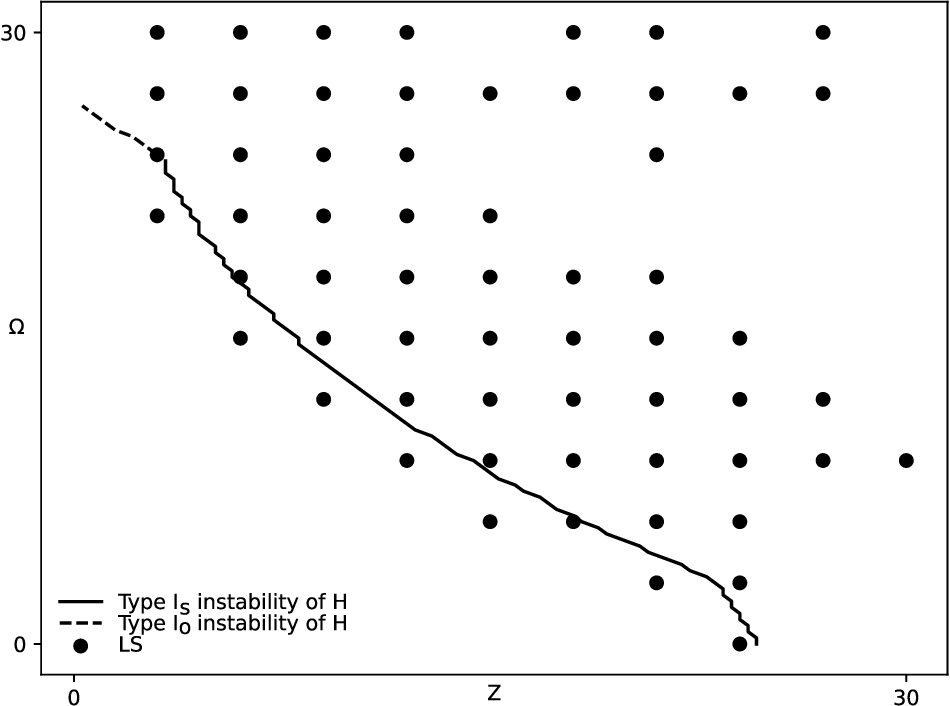}
		\caption{\label{fig:stability_diagram} Stability diagram obtained from linear stability analysis. The homogeneous steady state H is unstable above the solid and dotted lines. Parameter values as in Fig. 1 of the main text. }
	\end{figure}

\begin{figure}
		\includegraphics[width=\columnwidth]{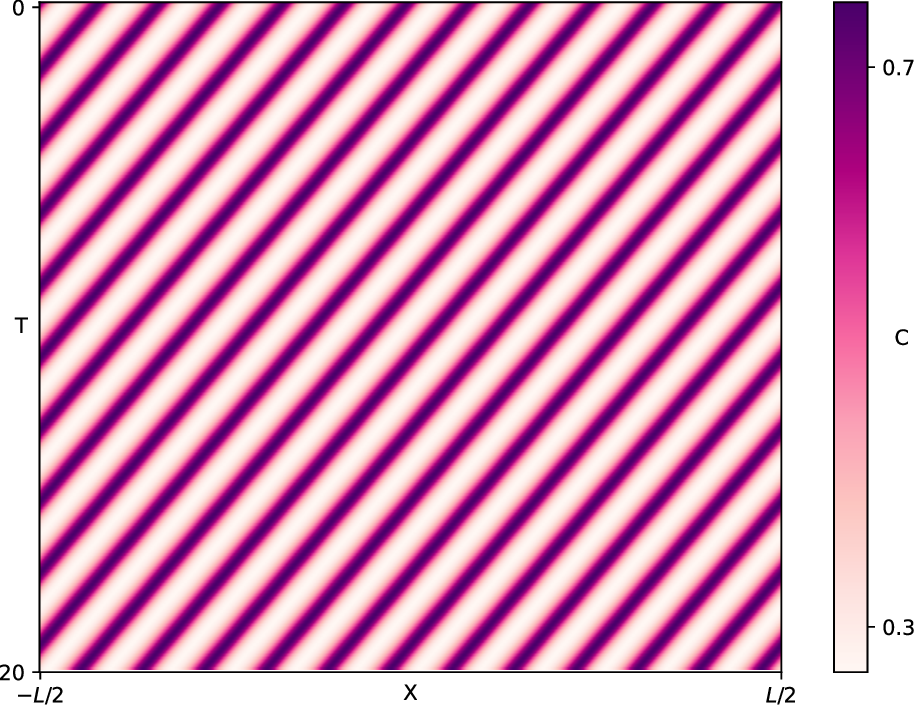}
		\caption{\label{fig:wave} Travelling wave state observed at $Z = 0$, $\Omega = 30$ (other parameters as in Fig. 1 of the main text). Initial condition is H perturbed by weak white noise.}
	\end{figure}

	We are ready to investigate the stability of H by numerically evaluating the eigenvalues $\Phi = \Phi(K)$ of $M$. In particular, we search for finte wavelength stationary (type $\mathrm{I_s}$) or oscillatory (type $\mathrm{I_o}$) instabilities. In SM Fig.~\ref{fig:stability_diagram}, we illustrate a cut in parameter space where both instabilities can take place. To calculate the stability boundaries, we used a custom Julia code that is freely available online (\url{https://github.com/lucabrb/Barberi-Kruse-2022}). In  SM Fig.~\ref{fig:stability_diagram}, we also indicate the presence of LSs. Note that instabilities are possible even if the fluid is passive ($Z=0$), in which case we observe travelling waves beyond the Type I$_o$ instability, SM Fig.~\ref{fig:wave}.
	
	The procedure used to locate the LSs in SM Fig.~\ref{fig:stability_diagram} is the following. We initialize the dynamical fields $A = C, N_a, N_b$ with a localized initial condition $A(X, T = 0) = A_\mathrm{H} [1 + S^0(X)]$, with $S^0(X) = S(X) - (1/L) \int_{-L/2}^{L/2} S(X) dX$, where $S(X) = \theta(X + 5)\theta(X - 5)$ and $\theta$ is the Heaviside step function. We then integrate the equations numerically, for a total time $T_\mathrm{tot} = 10^4$ and determine the presence of a static or oscillating LS depending on the result. We label the resulting state as localized if relative deviations from the homogeneous state reached at $X = \pm L/2$ are smaller than $10^{-3}$ in at least one eighth of the system. Note that, beyond the stability boundary, LSs are likely more widespread than shown in Fig.~\ref{fig:stability_diagram}. However, their observation requires either different initial conditions or longer runtime. Also note that some of the LSs highlighted in Fig.~\ref{fig:stability_diagram} are oscillatory. 
		
	\subsection{Inner wavelength of localized patterns}\label{sec:LSA_2}
	Localized states (LSs) feature a region of enhanced active fluid and nucleator density. If the local average nucleator density is sufficiently high, it can induce local pattern formation in this region, through instabilities like those discussed in Sec.~\ref{sec:LSA_1}. Indeed, both $Z$ and $\Omega$ increase with increasing average nucleator density $\bar{n}$, SM Table~\ref{table:parameters}. This mechanism is responsible for the emergence of localized patterns with specific inner wavelength, as we show below.
	
	To start, we define the local average nucleator density as $\bar{N}_\text{loc} = \frac{1}{X_r - X_l} \int_{-X_l}^{X_r} dX \left(N_a + N_i\right)$, where $X_l$ ($X_r$) is the first point from the left (right) where the relative deviation of $N_i$ from its homogeneous state at $X = -L/2$ ($X = L/2$) is larger than 10\%. We also define the width $W$ of the localized pattern as its Full Width Half Maximum (FWHM). The values of $\bar{N}_\text{loc}$ and $W$ for a set of localized patterns, LP$_{2-7}$, are provided in SM Table~\ref{table:lp}. In the same table, we also provide the inner pattern's wave number, $K_\text{obs}$, as observed in the simulations. We now compare this wave number with $K^*$, obtained from linear stability analysis of the homogeneous state with $\bar{N} = \bar{N}_\text{loc}$ and all other parameters as for the corresponding LP. Specifically, $K^*$ is the fastest growing mode of the linearized dynamics for a system of size $W$, implying $K = n\Delta K$, where $n \in \mathbb{N}$ and $\Delta K = \pi/W$. The list of $K^*$ for each LP is provided in SM Table~\ref{table:lp}. In SM Table~\ref{table:lp}, we also provide the interval of unstable modes around $K^*$, $K^* \in \left[K^*_\text{left}, K^*_\text{right}\right]$. 
	
	In the case of LP$_{6}$ and LP$_{7}$, the wave number of the fastest growing linear mode is very close to that of the observed pattern in the high-density region, $K^* \simeq K_\text{obs}$. In the case of LP$_{4}$ and LP$_{5}$, $K^*$ and $K_\text{obs}$ are rather different. Still, the 
	interval of unstable modes includes the observed wave number, $K_\text{obs} \in \left[K^*_\text{left}, K^*_\text{right}\right]$. Note that LP$_{4}$ and LP$_{5}$ coexist with LP$_{6}$, with which they also share similar values of $W$ and $\bar{N}_\text{loc}$. This suggests that the inner patterns of LP$_{4}$ and LP$_{5}$ might emerge from subcritical instabilities. In the case of LP$_{3}$, wave numbers agree up to an error $K^* - K_\text{obs} \simeq \Delta K$. Finally, the result for LP$_{2}$ is of similar quality as that for LP$_{4}$ and LP$_{5}$. The origin of the inaccuracy might be similar in the three cases. Indeed, LP$_{2}$ coexists with LP$_{3}$, with which it also shares similar values of $W$ and $\bar{N}_\text{loc}$, suggesting the inner pattern of LP$_{2}$ might emerge from a subcritical instability.
	
	\begin{table}[h!]
		\begin{center}
			\begin{tabular}{| c || c | c | c | c |} 
				\hline
				 & $\bar{N}_\text{loc}$ & W & $K_\text{obs}$ & $K^* \in \left[K^*_\text{left}, K^*_\text{right}\right]$ \\ 
				\hline\hline
				LP$_2$ & 1.75 & 7.56 & 2.47 & $4.15 \in [0.42, 9.56]$ \\ 
				\hline
				LP$_3$ & 1.68 & 7.93 & 3.14 & $3.57 \in [0.40, 8.32]$ \\
				\hline
				LP$_4$ & 1.58 & 11.19 & 2.79 & $4.21 \in [1.12, 9.27]$ \\
				\hline
				LP$_5$ & 1.55 & 11.50 & 3.26 & $4.10 \in [1.09, 8.74]$ \\
				\hline
				LP$_6$ & 1.54 & 11.50 & 3.81 & $3.83 \in [1.09, 8.47]$ \\
				\hline
				LP$_7$ & 1.50 & 12.60 & 3.97 & $3.99 \in [1.25, 8.5]$ \\
				\hline
			\end{tabular}
			\caption{Geometric features of six different localized patterns, LP$_{2-7}$. For definitions, see Sec.~\ref{sec:LSA_2}. LP$_{2-7}$ are the same as in Fig. 1 of the main text.}
			\label{table:lp}
		\end{center}
	\end{table}
	
	\section{Spatial dynamics}
	In the spatial dynamics framework, Eqs.~\eqref{eq:actin_dynamics}--\eqref{eq:nucleator_dynamics_b} map onto the following system of ordinary differential equations (ODEs):
	\begin{align} 
			\partial_X C&= P_C, \label{eq:spatial_dynamics_first}\\
			\partial_X N_a&= P_a, \\
			\partial_X N_i&= P_i, \\
			\partial_X V&= P_V, \\
			\partial_X P_C&= \frac{1}{\mathscr{D}_c}\left(V P_C + C P_V - A N_a + k_d C\right), \\
			\partial_X P_a&= \frac{1}{\mathscr{D}_a}\left[V P_a + N_a P_V -\Omega_0 \left(1 + \Omega N_a^2\right)N_i + \left(\Omega_{d,0} + \Omega_d C\right) N_a\right], \\
			\partial_X P_i&= V P_i + N_i P_V +\Omega_0 \left(1 + \Omega N_a^2\right)N_i - \left(\Omega_{d,0} + \Omega_d C\right) N_a, \\
			\partial_X P_V&= V - 2 Z C P_C + 3 B C^2 P_C. \label{eq:spatial_dynamics_last}
	\end{align}
	The mapping above is mathematically exact. Therefore, the existence of steady states in Eqs.~\eqref{eq:actin_dynamics}--\eqref{eq:nucleator_dynamics_b} implies the existence of specific solutions of the ODE system. Periodic boundaries imply that all solutions of the ODE system are closed orbits, with the exception of homogeneous steady states, which map onto fixed points.

The eigenvalues in Fig.~3c of the main text are calculated by performing a linear stability analysis of Eqs.~\eqref{eq:spatial_dynamics_first}--\eqref{eq:spatial_dynamics_last} around the homogeneous solution H. 
	
	\section{Other supplementary tables and figures}
	
	\begin{table}[h!]
		\begin{center}
			\begin{tabular}{| c | c | c |} 
				\hline
				Parameter & Expression & Value \\
				\hline\hline
				$L$ & $l/\lambda$ & $10\pi$ \\ 
				\hline
				$\mathscr{D}_c$ & $D_c \tau/\lambda^2$ & 0.01 \\
				\hline
				$\mathscr{D}_a$ & $D_a \tau/\lambda^2$ & 0.1 \\
				\hline
				$\mathscr{D}_i$ & $D_i \tau/\lambda^2$ & 1 \\
				\hline
				$A$ & $A\tau$ & 1 \\
				\hline
				$K_d$ & $k_d\tau$ & 1 \\
				\hline
				$\Omega_0$ & $\omega_0 \tau$& 0.5 \\
				\hline
				$\Omega$ & $\omega \bar{n}^2$ & see text \\
				\hline
				$\Omega_{d,0}$ & $\omega_{d,0} \tau$ & see text \\
				\hline
				$\Omega_d$ & $\omega_d \bar{n} \tau$ & see text \\
				\hline
				Z & $(\zeta\Delta\mu)_0 \overline{n}^2\tau/\gamma\lambda^2$ & see text \\
				\hline
				B & $b \bar{n}^3 \tau/\gamma\lambda^2$ &Z \\
				\hline
			\end{tabular}
		\caption{List of non-dimensional parameters.}
		\label{table:parameters}
		\end{center}
	\end{table}
	
	\begin{figure}
		\includegraphics[width=\columnwidth]{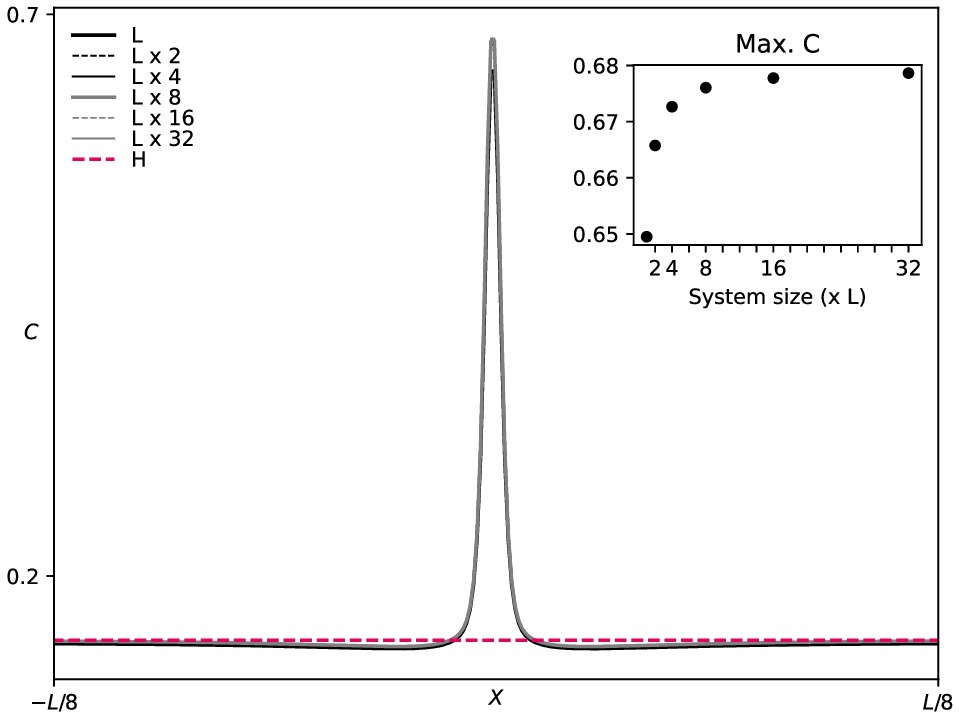}
		\caption{\label{fig:slocalizedness} Actomyosin profile, $C$, of the LS in Fig. 3a of the main text, at varying system size ($L \times 1, 2, 4, 8, 16, 32$) and all other parameters kept constant (in particular, $\bar{N} = 1$). The pink dashed line represents the homogeneous steady state, $H$. To better visualize the LS profile, only a portion $X \in [-L/8, L/8]$ of the horizontal axis is shown. In simulations, the $X$ axis is discretized such that grid spacing $\Delta X = L / 1024$ is independent of system size. (Inset) The maximum of $C$, \emph{i.e.} $C(X=0)$, saturates to $C_\text{max} \simeq 0.68$ with increasing system size.}
	\end{figure}

	\begin{figure}
		\includegraphics[width=\columnwidth]{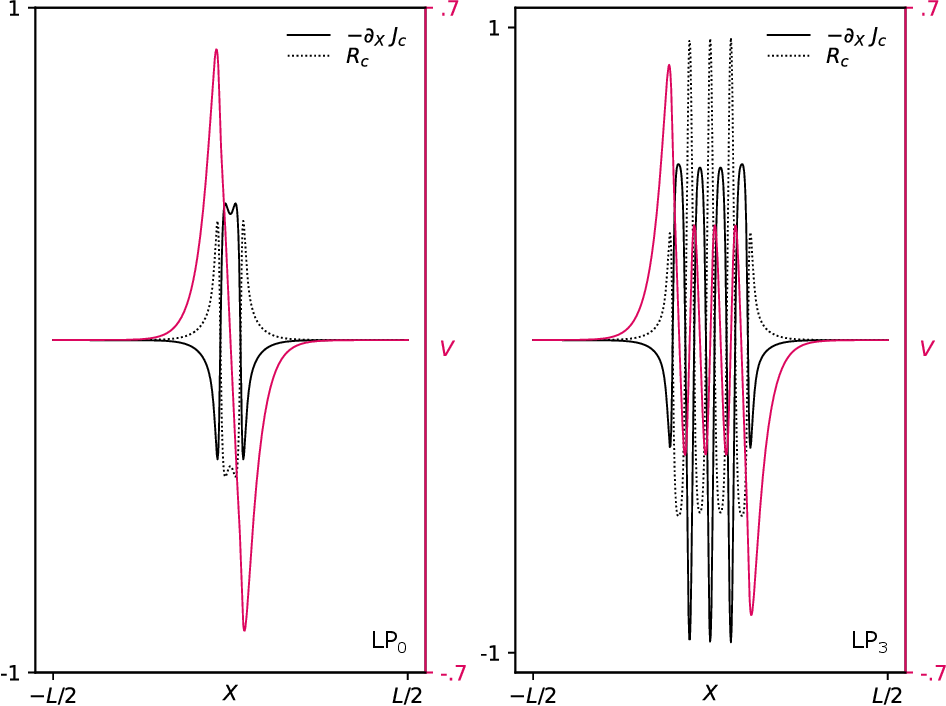}
		\caption{\label{fig:transport_vs_chemistry} This figure corresponds to the localized patterns LP$_0$ (left) and LP$_3$ (right) in Fig. 1(a, c) of the main text. $J_c$ is defined as in Sec.~\ref{sec:LSA}, while $R_c$ is the right hand side of Eq.~\eqref{eq:actin_dynamics}.}
	\end{figure}

	\begin{figure}
		\includegraphics[width=\columnwidth]{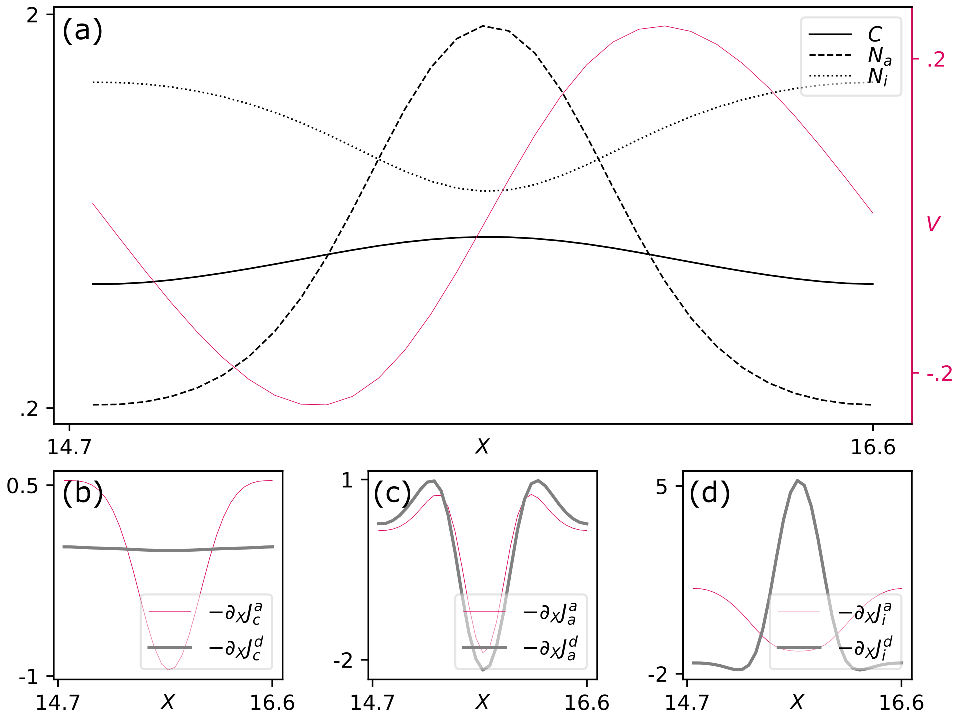}
		\caption{\label{fig:inner_peak_eq} This figure zooms into the central inner peak of LP$_3$ in Fig. 2c of the main text. (a) Density (thick black lines) and velocity (thin pink line) profiles at the peak. (b) Advective, $-\partial_X J^a_c = -\partial_X CV$, and diffusive, $-\partial_X J^d_c = D_c \partial_X^2 C$, contribution to the transport term, $-\partial_X J_c = -\partial_X (J_c^a + J_c^d)$, of Eq.~\eqref{eq:actin_dynamics}. (c, d) Same as in panel (b), for the active (c) and inactive (d) nucleators.}
	\end{figure}

	\begin{figure}
		\includegraphics[width=\columnwidth]{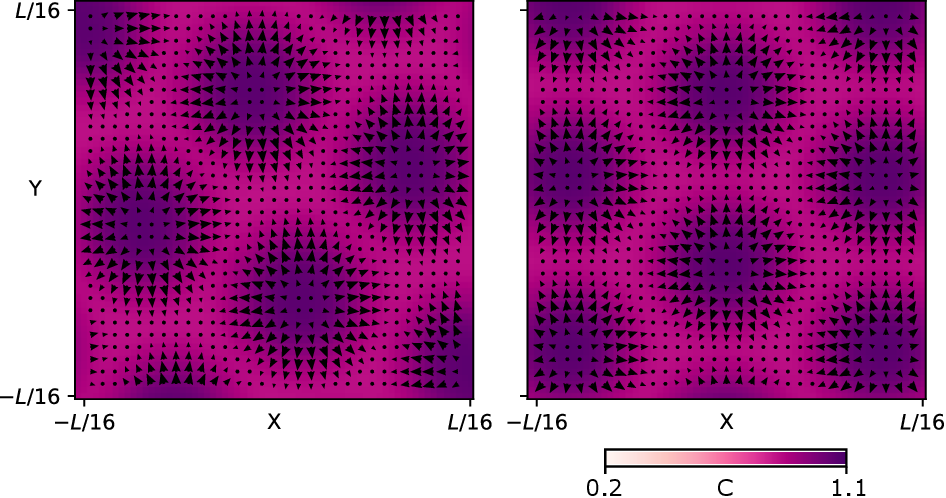}
		\caption{\label{fig:2d_outflow} This figure corresponds to Fig. 4 of the main text. (left) Zoom-in of Fig. 4b of the main text. (right) Zoom-in of Fig. 4d of the main text. Black arrowheads represent the velocity field. Left and right panel have the same vertical axis.}
	\end{figure}
	